\documentclass[aps,longbibliography, nofootinbib,eqsecnum,twocolumn]{revtex4-1}
\usepackage{graphicx}
\usepackage{stmaryrd}
\usepackage{amsmath, amsfonts, mathrsfs, braket}
\usepackage{color}
\usepackage{xspace}
\usepackage{comment}
\usepackage{bm}
\definecolor{lightblue}{rgb}{0.13, 0.26, 0.99}

\usepackage[
colorlinks=true,
urlcolor=blue,
citecolor=blue,
linkcolor=blue,
hyperfootnotes=false]{hyperref}

\newcommand{\im}{\operatorname{Im}}

\allowdisplaybreaks

\begin{document}

\title{Quantized edge magnetizations and their symmetry protection \\ in one-dimensional quantum spin systems}
\author{Shunsuke C. Furuya}
\author{Masahiro Sato}
\affiliation{Department of Physics, Ibaraki University, Mito, Ibaraki 310-8512, Japan}
\date{\today}

\begin{abstract}
The bulk electric polarization works as a nonlocal order parameter that characterizes topological quantum matters.
Motivated by a recent paper [H. Watanabe \textit{et al.}, Phys. Rev. B {\bf 103}, 134430 (2021)],
we discuss magnetic analogs of the bulk polarization in one-dimensional quantum spin systems, that is, quantized magnetizations on the edges of one-dimensional quantum spin systems.
The edge magnetization shares the topological origin with the fractional edge state of the topological odd-spin Haldane phases.
Despite this topological origin, the edge magnetization can also appear in topologically trivial quantum phases.
We develop straightforward field theoretical arguments that explain the characteristic properties of the edge magnetization.
The field theory shows that a U(1) spin-rotation symmetry and a site-centered or bond-centered inversion symmetry protect the quantization of the edge magnetization.
We proceed to discussions that quantum phases on nonzero magnetization plateaus can also have the quantized edge magnetization that deviates from the magnetization density in bulk.
We demonstrate that the quantized edge magnetization distinguishes two quantum phases on a magnetization plateau separated by a quantum critical point.
The edge magnetization exhibits an abrupt stepwise change from zero to $1/2$ at the quantum critical point because the quantum phase transition occurs in the presence of the symmetries protecting the quantization of the edge magnetization.
We also show that the quantized edge magnetization can result from the spontaneous ferrimagnetic order.
\end{abstract}

\maketitle

\section{Introduction}\label{sec:intro}

The electric polarization is a fundamental object of electromagnetism~\cite{landau_continuous_media}.
The polarization also receives attention to its close connection with topological quantum matters.
Even though intrinsic to the bulk, the electric polarization manifests itself as an accumulation of charges on the surface~\cite{vanderbilt_polarization_1993}.
Surface electric charges work as topological indices of topological insulators and higher-order ones~\cite{benalcazar_multipole_2017,Benalcazar2017_hoti,qi_tft_polarization,ezawa_hoti}.
Importantly, we can experimentally access the surface charge more easily than the entanglement entropies and spectra~\cite{pollmann_haldane_2010,pollmann_haldane_2012} and other related nonlocal order parameters~\cite{aklt_comm_math,Kennedy_Z2Z2,Kohmoto_Z2Z2,nakamura-todo}.

Considering the fruitful relations between the surface charges and topological quantum matters, we can expect that it holds promise to consider a magnetic analog of the surface electric charge.
A leading candidate is a surface magnetization because the magnetization is a conserved quantity associated with a global U(1) symmetry similarly to the electric charge.
Recently, Watanabe \textit{et al.}~\cite{watanabe_corner} 
found fractionally quantized magnetizations, $M^z=\pm S/2^d$, accumulated on edges ($d=1$) or corners ($d=2,3$) of $d$-dimensional cubic-lattice Heisenberg antiferromagnetic (HAFM) models under a staggered magnetic field.
They showed that these fractional edge and corner magnetizations qualify as the magnetic analog of the bulk polarization.

We can expect that the edge or corner magnetization will detect some topological properties of quantum magnets.
Reference~\cite{watanabe_corner} suggested that the edge magnetization $M^z=\pm 1/2$ in the spin-1 chain is ``reminiscent of the $S=1/2$ edge mode'' of the spin-1 Haldane phase.
On the other hand, the edge magnetization also emerges in topologically trivial phases such as the forced N\'eel phase of the spin-$1/2$ chain.

However, it is still puzzling what kind of topological property the edge magnetization detects.
The two spin chains mentioned above seem to have different topological profiles.
The topological aspect of the edge magnetization as the bulk polarization is nontrivial even in one-dimensional systems, which is worth further investigations, in particular, from a field-theoretical point of view.
One-dimensional quantum theories will be useful to build two- or three-dimensional quantum states with quantized corner magnetizations~\cite{watanabe_corner} similarly to coupled-wire constructions of topological states~\cite{kane_cwc_2002,kane_cwc_2014,meng_cwc_csl,lecheminant_cwc_na_csl}.

This paper develops a quantum field theory of the edge magnetization as the magnetic analog of the bulk electric polarization.
We make twofold claims.
One is that the quantum field theory clarifies that the symmetry-protected edge state of the topological Haldane phase and the quantized edge magnetization of the topologically trivial phase have the same origin.
They originate from a zero mode of a nonlocal boson field whose spatial gradient represents the magnetization density.
The other is that the edge magnetization also appears in quantum spin systems in uniform magnetic fields.
The staggered magnetic field is not always necessary to induce the quantized edge magnetization.
The quantized edge magnetization can be induced by spatially modulated exchange interactions and even by spatially uniform interactions.

We organize this paper to make it accessible to a broad readership without going deeply into technical details of quantum field theories, while details are given in Appendices to make the paper self-contained.
We first give a field-theoretical interpretation of the quantized edge magnetization in the general spin-$S$ HAFM chains (Sec.~\ref{sec:zero_field}), where we see the physical meaning of the Gaussian convolution introduced in Ref.~\cite{watanabe_corner}.
Next, we apply the magnetic field to quantum spin systems.
Section~\ref{sec:plateau} deals with a $1/2$ magnetization plateau of the spin ladder, where we arrive at a generalized definition of the edge magnetization valid on the magnetization plateaus.
Sections~\ref{sec:J4} and \ref{sec:ferri} also discuss edge magnetizations on the magnetization plateaus, but their ground states are much more nontrivial than that dealt with in Sec.~\ref{sec:plateau}.
We will see an interesting topological quantum phase transition on the 1/2 magnetization plateau accompanied by an abrupt change of the edge magnetizations and the total charge in Sec.~\ref{sec:J4}.
Section~\ref{sec:ferri} discusses the ferrimagnetic ground state with the edge magnetization.
We summarize the paper in Sec.~\ref{sec:conclusions}.
Appendices describe details of quantum field theories.
Note that Appendices~\ref{app:m=0} and \ref{app:m>0} contain novel results such as what we call ``semiclassical bosonization formulas''.

\section{Unification of  topological edge state and edge magnetization}\label{sec:zero_field}

\subsection{Quantized edge magnetization in topologically trivial phase of spin-1/2 chain}\label{sec:zero_field_1/2}

We start our discussions by field-theoretically interpreting results of Ref.~\cite{watanabe_corner} for one-dimensional systems.
The simplest situation is the spin-$1/2$ HAFM chain with the staggered magnetic field.
The Hamiltonian with the open boundary condition (OBC) is given by
\begin{align}
    \mathcal{H}=J\sum_{j=1}^{L-1} \bm{S}_j\cdot\bm{S}_{j+1}+h_s\sum_{j=1}^L (-1)^jS_j^z,
    \label{H_chain}
\end{align}
where $\bm S_j$ is the spin-$1/2$ operator at the $j$th site and $J>0$ is the antiferromagnetic exchange coupling and $h_s>0$ is the staggered magnetic field.
According to the Marshall-Lieb-Mattis theorem~\cite{marshall,lieb-mattis}, the ground state of the model \eqref{H_chain} satisfies
\begin{align}
    \sum_{j=1}^L\braket{S_j^z}=0.
    \label{neutral_Sz}
\end{align}
If we regard $S_j^z$ as a charge, Eq.~\eqref{neutral_Sz} gives a charge neutrality condition.
Precisely speaking, the neutrality condition \eqref{neutral_Sz} is met only when $L$ is even.
We assume even $L$ throughout this paper.

The low-energy physics of this spin chain is described by a quantum field theory of an interacting U(1) compactified boson $\phi$ with a trigonometric potential, called the sine-Gordon theory~\cite{giamarchi_book,gogolin_textbook}:
\begin{align}
    \mathcal H_s &=\int_0^L dx \biggl[ \frac v{2\pi K}(\partial_\mu\phi)^2 +g_s\sin(2\phi)\biggr].
    \label{H_SG}
\end{align}
Hereafter, we employ a unit system with $\hbar=a_0=1$ ($a_0$ is the lattice spacing) for simplicity, but will call $a_0$ back whenever we need.
The first term of Eq.~\eqref{H_SG} is the kinetic term, where
$(\partial_\mu\phi)^2=v^{-2} (\partial_\tau\phi)^2+(\partial_x\phi)^2$ and $x$ and $\tau$ are the space and the imaginary time, respectively.
This paper mainly uses the imaginary-time formalism for later convenience.
The coupling of the sine potential is proportional to the staggered field, $g_s\propto h_s$ .
The parameter $v$ denotes the velocity of the boson field, $\phi$~\cite{giamarchi_book}.
For the spin-1/2 Heisenberg chain, $v=\pi J/2$~\cite{giamarchi_book}.
The $\phi$ field is related to the spin operator $\bm{S}_j$ as~\cite{giamarchi_book,gogolin_textbook}
\begin{align}
    \bm{S}_j=\bm{M}_j+(-1)^j\bm{N}_j
    \label{bosonization}
\end{align}
where the $z$ component is given by 
\begin{align}
    M_j^z=\frac{a_0}{\pi}\partial_x\phi,\qquad
    N_j^z=a_1\sin(2\phi),
    \label{MzNz}
\end{align}
with a constant $a_1$~\cite{hikihara_coeff}.
If $h_s=0$, the ground state of the spin-1/2 HAFM chain \eqref{H_chain} is the gapless Tomonaga-Luttinger (TL) liquid state~\cite{giamarchi_book,gogolin_textbook} governed by the kinetic term, $v(\partial_\mu\phi)^2/2\pi K$.
The sine potential, $\sin(2\phi)$, favoring a constant $\phi$ competes with the kinetic term favoring a constant $\partial_\mu\phi$.
The staggered magnetic field $h_s$ yields the spin gap by locking the $\phi$ field to a minimum of the sine potential, 
\begin{align}
	\bar \phi = - \frac{\pi}{4} \mod \pi.
	\label{lock_hs}
\end{align}

In one-dimensional quantum many-body systems, the bulk polarization $\mathcal{P}$ is precisely given by the Resta's formula~\cite{resta_p,watanabe_polarization},
\begin{align}
    \mathcal{P} &= \frac{1}{2\pi}\im\ln \braket{U}.
    \label{P_resta}
\end{align}
Resta gave $U=\exp(i\frac{2\pi}L \sum_{j=1}^L j n_j)$ with the one-particle density $n_j$~\cite{resta_p}.
In quantum spin systems, $S_j^z$ plays the role of $n_j$~\cite{giamarchi_book, watanabe_corner}.
Hence, we here define $U$ as
\begin{align}
    U &= \exp\biggl( i\frac{2\pi}{L}\sum_{j=1}^L j S_j^z \biggr),
    \label{U_def}
\end{align}
in the model \eqref{H_chain}.
The nonlocal operator \eqref{U_def} is deeply related to the Lieb-Schultz-Mattis (LSM) theorem~\cite{lsm,oshikawa_lsm_flux,oya,yao_boundary} that allows for classification of gapless quantum phases~\cite{cho_anomaly,furuya_wzw,yao_su_n}.
The ground state expectation value $\braket{U}$ is called the polarization amplitude and was previously studied in valence-bond-solid phases~\cite{nakamura-todo} and in the TL-liquid phase~\cite{kobayashi_pol_amp,nakamura_pol_amp,furuya_pol_amp}.

The bosonization formula $M_j^z=\partial_x\phi/\pi$ gives~\cite{furuya_pol_amp}
\begin{align}
    U &= \exp[2i(\phi(L)-\bar\phi)],
    \label{U_phi} \\
    \bar\phi &= \frac 1L \int_0^L dx \, \phi(x),
    \label{bar_phi}
\end{align}
where $\bar\phi$ is the zero mode of the $\phi$ boson~\cite{giamarchi_book}.
$\bar\phi$ can also be seen as the spatial average of $\phi$.
The representation \eqref{U_phi} implies that $\phi(L)$ at the right edge of the chain deviates from the average $\bar\phi$ when $\mathcal{P}\not=0$.
The field $\phi(x=0)$ at the left edge also deviates from $\bar\phi$ because the OBC on the spin chain imposes the following boundary condition on $\phi(x)$~\cite{eggert_obc} (Appendix~\ref{app:obc}):
\begin{align}
    \phi(0) = \phi(L) = 0 \mod \pi.
    \label{bc_phi}
\end{align}
The lockings \eqref{bc_phi} of $\phi$ at the edges hold irrespective of the locking~\eqref{lock_hs} in the bulk.
The boundary condition \eqref{bc_phi} holds even when the bulk is gapless.
The boundary condition \eqref{bc_phi} is automatically consistent with the charge neutrality because $\sum_{j=1}^L\braket{S_j^z}=\braket{(\phi(L)-\phi(0))}/\pi=0$.
Equation~\eqref{bc_phi} also indicates that $\bar\phi$ generally represents the locking position of $\phi$ in the bulk.
With the OBC, the polarization amplitude $\braket{U}$ is thus given by
\begin{align}
    \braket{U}=\braket{e^{-2i\bar\phi}} \approx e^{-2i\braket{\bar\phi}}.
\end{align}
Note that the latter approximate equality holds when a relevant interaction strongly locks $\phi$ in bulk to a constant.
The approximation becomes more accurate when the bulk excitation gap due to the staggered field becomes larger.
We thus find that the bulk polarization \eqref{P_resta} is given by the zero mode,
\begin{align}
    \mathcal{P}\approx -\frac{1}{\pi} \braket{\bar\phi} \mod 1,
    \label{P_phi}
\end{align}
since $\phi$ is real [see Eq.~\eqref{MzNz}].

The staggered magnetic field leads to the locking \eqref{lock_hs} of $\bar\phi$.
Hence, the bulk polarization $\mathcal{P}=1/4\mod1$ follows.
The locking value $\phi(L)=0$ at the right edge $x=L$ indeed deviates from the bulk one~\eqref{lock_hs}.

\begin{figure}[t!]
    \centering
    \includegraphics[bb = 0 0 900 300, width=\linewidth]{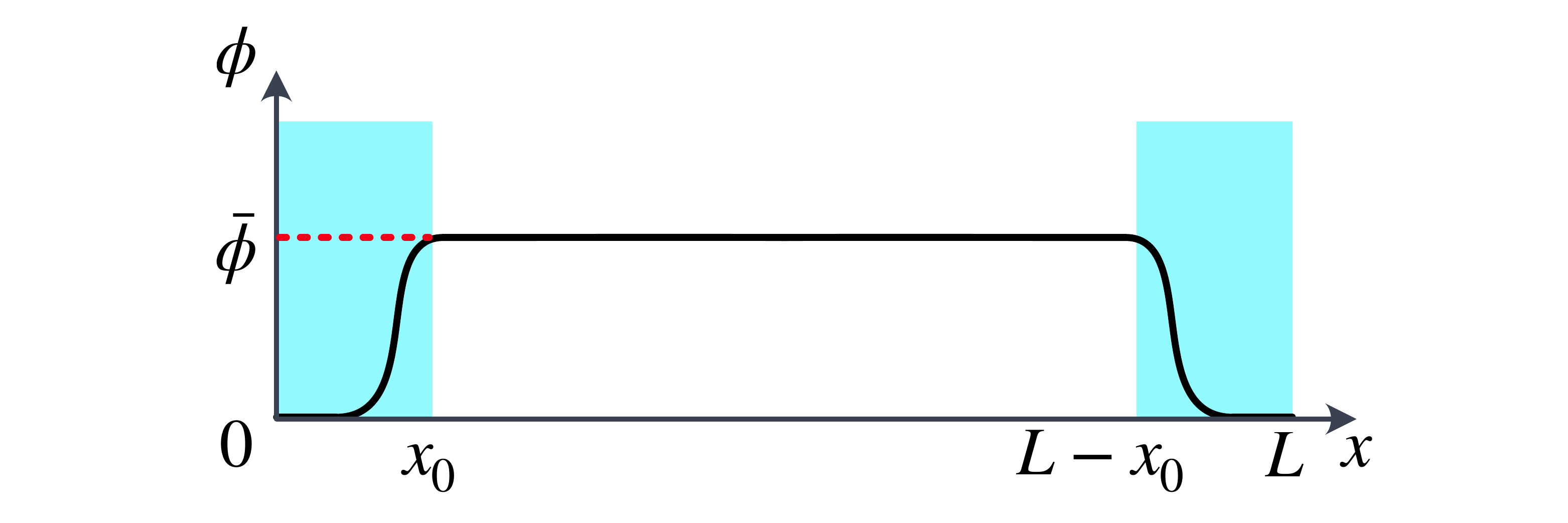}
    \caption{Schematic $x$ dependence of $\phi(x)$ of sine-Gordon model \eqref{H_SG}.
        The shaded areas depict the left and right edges, $\mathcal L$ and $\mathcal R$.
    }
    \label{fig:phi}
\end{figure}

We can relate the bulk polarization $\mathcal P$ and the edge magnetization introduced by Ref.~\cite{watanabe_corner} as follows.
It is generally not obvious how to clearly distinguish the bulk from the edge.
There  is no clear border between the bulk and the edge.
However, when $\bar\phi\not=0 \mod \pi$,  we can distinguish the bulk and the edge by setting a sufficiently small cutoff $\epsilon>0$.
Let us define the bulk part, $\mathcal B$, of the spin chain as a region where $\phi(x)$ is locked to $\bar\phi$,
\begin{align}
    \mathcal B &= \{x \in [0,L] \,| \,0\le   |\phi(x)-\bar\phi| < \epsilon \mod \pi \}.
    \label{B_def}
\end{align}
It is convenient to split $\mathcal B = \mathcal B_l \cup \mathcal B_r$ into two:
\begin{align}
    \mathcal B_l 
    &= \{x \in [0,L/2]\, | \, 0\le | \phi(x)-\bar\phi| < \epsilon \mod \pi\}, \\
    \mathcal B_r
    &= \{ x\in (L/2,L]\, |\,  0\le  |\phi(x)-\bar\phi| < \epsilon \mod \pi \}.
\end{align}
The left edge $\mathcal L$ and the right edge $\mathcal R$ are then defined as
\begin{align}
    \mathcal L &= [0,\, L/2] \setminus \mathcal B_l, \\
    \mathcal R &= (L/2,\, L]\setminus \mathcal B_r,
\end{align}
where $\setminus$ denotes the set difference.
We can rewrite $\mathcal L = [0, x_0)$ and $\mathcal R = (L-x_0, L]$ by using the smallest $x_0 \in \mathcal B$  (the shaded areas of Fig.~\ref{fig:phi}).
Then, the uniform part, $M_j^z$, of $S_j^z$ on the left edge $\mathcal L$ corresponds to $-\mathcal P$ because
\begin{align}
    \sum_{ja_0\in \mathcal L} \braket{M_j^z}
    &= \int_{x\in \mathcal L} dx \, \frac{1}{\pi} \braket{\partial_x\phi(x)}
    \notag \\
    &= \frac{1}{\pi} \braket{[ \phi(x_0)-\phi(0)]}
    \notag \\
    &\approx- \mathcal P+O(\epsilon).
    \label{ML}
\end{align}
Likewise, we obtain on the right edge,
\begin{align}
    \sum_{ja_0 \in \mathcal R} \braket{M_j^z} 
    &= \int_{x\in\mathcal R} dx \, \frac{1}{\pi} \braket{\partial_x\phi(x)}
    \notag \\
    &= \frac{1}{\pi} \braket{[\phi(L)-\phi(L-x_0)]}
    \notag \\
    &\approx \mathcal P + O(\epsilon).
    \label{MR}
\end{align}
Hence, we find that Eqs~\eqref{ML} and \eqref{MR} equal to the quantized edge magnetizations reported in Ref.~\cite{watanabe_corner}.

To further support this claim, we recall the definition of the edge magnetization by Watanabe \textit{et al.}~\cite{watanabe_corner}.
They first convoluted a Gaussian function,
\begin{align}
    g(r)= \frac{1}{\sqrt{2\pi\lambda^2}}\exp\biggl(-\frac{r^2}{2\lambda^2}\biggr),
\end{align}
with $\lambda>0$ to the magnetization $\braket{S_r^z}$:
\begin{align}
    m^z(r)=\sum_{r'=1}^Lg(r-r')\braket{S_{r'}^z}.
    \label{mz_conv}
\end{align}
Next, they defined the edge magnetizations on both  edges by using the convoluted magnetization \eqref{mz_conv}.
\begin{align}
     M_{\rm left}^z = \int_{-\infty}^{\frac{L+1}2} dr \, m^z(r), \quad M_{\rm right}^z = \int_{\frac{L+1}2}^\infty dr \, m^z(r).
     \label{Ml_Mr}
\end{align}
The rapid oscillation $(-1)^j$ over the finite Gaussian window suppresses the staggered component $N_j^z$'s contribution to $m^z(r)$.
We can rewrite $M_{\rm left}^z$ as
\begin{align}
	M_{\rm left}^z 
	&= \int_0^L dr'\, \frac 1\pi \braket{\partial_{r'}\phi(r')} \int_{-\infty}^{\frac{L+1}{2}} dr \, g(r-r').
\end{align}
The integral about $r$ approximately gives
\begin{align}
	\int_{-\infty}^{\frac{L+1}{2}} dr \, g(r-r') &\approx \left\{
	\begin{array}{ccc}
		1 & & (\text{for } r'\lesssim \tfrac{L+1}2+\lambda) \\
		&& \\
		0 & & (\text{otherwise})
	\end{array}
	\right.
\end{align}
Therefore, $M_{\rm left}^z$ is reduced to
\begin{align}
	M_{\rm left}^z &\approx \int_{-\infty}^{\frac{L+1}{2} +\lambda} dr' \, \frac 1\pi \braket{\partial_{r'} \phi(r')}
	\notag \\
	&= \frac 1\pi \biggl\langle \biggl(\phi\biggl(\frac{L+1}2 +\lambda \biggr) - \phi(0) \biggr)\biggr\rangle
	\notag \\
	&= \frac 1\pi \bar\phi
	\notag \\
	&\approx -\mathcal P.
\end{align}
By applying a similar procedure to $M_{\rm right}^z$, we obtain
\begin{align}
    M_{\rm left}^z &\approx \sum_{ja_0\in \mathcal L} \braket{M_j^z} \approx - \mathcal P, 
    \label{Mz_left} \\
    M_{\rm right}^z &\approx \sum_{ja_0\in \mathcal R} \braket{M_j^z} \approx \mathcal P,
    \label{Mz_right}
\end{align}
We can employ any other smooth normalized window function $g(r)$ (e.g. Lorentzian).
Regardless of the details or choice of $g(r)$, the convolution \eqref{mz_conv} with the finite window function $g(r)$ discards the staggered part of $S_j^z$.
Equations~\eqref{Mz_left} and \eqref{Mz_right} become more accurate as the lowest excitation gap in the bulk becomes larger.
Indeed, Ref.~\cite{watanabe_corner} demonstrated that the edge magnetizations $M_{\rm left}^z$ and $M_{\rm right}^z$ rapidly approach $-\mathcal P$ and $+\mathcal P$ respectively as the bulk excitation gap grows.

\subsection{Topological edge state and quantized edge magnetization in topological phase of spin-1 chain}

The sine-Gordon argument also applies to the spin-$1$ HAFM chain.
When the topological properties are concerned, we may identify the spin-$1$ HAFM chain as a two-leg spin-$1/2$ ladder with a weak ferromagnetic interchain interaction~\cite{schulz_spin-S,kim_ladder_haldane,hijii_ladder}, 
\begin{align}
    \mathcal H_{\rm ladder} &= J\sum_{j=1}^{L-1} \sum_{n=1,2} \bm S_{j,n} \cdot \bm S_{j+1,n}
    + J_\perp \sum_{j=1}^L \bm S_{j,1} \cdot \bm S_{j,2},
\end{align}
where each leg labeled by $n=1,2$ is the spin-$1/2$ HAFM chain and corresponds to the TL liquid of a boson $\phi_n$.
It is numerically confirmed that the weak ferromagnetic rung region ($-J_\perp/J\ll 1$) is adiabatically connected to the strong rung limit $-J_\perp/J\to -\infty$, that is, the spin-1 HAFM chain~\cite{hijii_ladder}.

The low-energy effective Hamiltonian is again the sine-Gordon theory \eqref{H_SG} of a boson, $\phi=\phi_1+\phi_2$~\cite{shelton_ladder,chitra-giamarchi},
\begin{align}
    \mathcal H_{\mathrm{c}} &=\int_0^L dx \biggl[ \frac v{2\pi K} (\partial_\mu\phi)^2 +g_c  \cos(2\phi)\biggr].
    \label{H_SG_cos}
\end{align}
This time, the trigonometric potential appears even in the absence of the staggered field.
The coupling $g_c\propto -J_\perp $ is proportional to the ferromagnetic rung interaction, $J_\perp<0$~\cite{shelton_ladder}.
If we impose the charge neutrality condition,
\begin{align}
    \sum_{j=1}^L \sum_{n=1,2} \braket{S_{j,n}^z}=0,
    \label{neutral_spin-1}
\end{align}
on the $\phi$ field, we obtain the OBC,
\begin{align}
    \phi(0)=\phi(L)=0\mod \pi,
    \label{OBC_spin-1}
\end{align}
in analogy with the spin-$1/2$ chain.

The ferromagnetic rung interaction locks $\bar\phi=\pm\pi/2\mod\pi$, leading to $\mathcal P = \mp 1/2 \mod 1$.
On the other hand, the antiferromagnetic coupling $J_\perp>0$ (i.e. $g_c<0$) locks $\bar\phi=0 \mod \pi$, leading to $\mathcal P = 0 \mod 1$.
The bulk polarization $\mathcal P$ thus distinguishes the topological Haldane phase for $J_\perp<0$ and the topologically trivial (rung-singlet) phase for $J_\perp>0$.

Similarly to the spin-$1/2$ chain with the staggered field, the spin-$1$ HAFM chain is accompanied by the bulk polarization $\mathcal{P}=\mp1/2$ that corresponds to the spin-$1/2$ edge state of the symmetry-protected topological spin-$1$ Haldane phase.
Unlike the spin-$1/2$ chain, the edge state is degenerate, and the degeneracy is protected by a symmetry such as a bond-centered inversion symmetry~\cite{pollmann_haldane_2010,pollmann_haldane_2012}.
The sine-Gordon theory \eqref{H_SG_cos} indicates that the ground state is doubly degenerate, $\ket{\rm GS_\pm}=(\ket{\psi_+} \pm \ket{\psi_-})/\sqrt{2}$.
Here, $\ket{\psi_\pm}$ denote bulk quantum states accompanied by edge magnetizations $(M_{\rm left}^z,\, M_{\rm right}^z)=(\mp 1/2, \, \pm 1/2)$, respectively.
In other words, $\ket{\psi_\pm}$ corresponds to a state with $\bar\phi=\mp \pi/2$, respectively.
We regard their superpositions, $\ket{\rm GS_\pm}$, as the degenerate ground states because $\ket{\psi_\pm}$ are not eigenstates of a bond-centered inversion $\mathcal I_b:\bm S_j \to \bm S_{L+1-j}$ but $\ket{\rm GS_\pm}$ are.
The bond-centered inversion acts on $\phi_1$ and $\phi_2$ as
\begin{align}
    \mathcal I_b \phi_n(x) \mathcal I_b^{-1} &= - \phi_n(L-x)  \mod \pi.
\end{align}
Accordingly, $\phi(x)=\phi_1(x)+\phi_2(x)$ transforms as
\begin{align}
    \mathcal I_b \phi(x) \mathcal I_b^{-1} &= - \phi(L-x)  \mod \pi.
\end{align}
Since $\bar\phi$ represents the locking position of $\phi$ in the bulk, $\mathcal I_b$ transforms $\bar\phi = \pi/2$ to $\bar\phi=-\pi/2 \mod \pi$  and vice versa.
Hence, it follows that $\mathcal I_b\ket{\psi_\pm}=\ket{\psi_\mp}$ and $\mathcal I_b \ket{\rm GS_\pm}=\pm \ket{\rm GS_\pm}$.
Whereas $\ket{\psi_\pm}$ is accompanied by the nonzero quantized magnetization, $\ket{\rm GS_\pm}$ are not.

Note that we found twofold degeneracy, not the fourfold one.
The ground state in the spin-1 Haldane phase with the OBC is fourfold degenerate because each end of the spin chain hosts the fractionalized $S=1/2$ spin.
This difference in the ground-state degeneracy originates from the charge neutrality condition \eqref{neutral_spin-1}.
Among the four ground states of the spin-1 HAFM chain with the OBC, two ground states live in the charge-neutral sector \eqref{neutral_spin-1} but the other two live out of it.
The sine-Gordon theory \eqref{H_SG_cos} formulated on the basis of the charge neutrality condition \eqref{neutral_spin-1} thus  predicts the twofold degeneracy.
The sine-Gordon theory gives the other two ground states of the spin-1 Haldane phase if we impose boundary conditions, $(\phi(0),\phi(L))=(0,\pi)$ or $(\phi(0),\phi(L))=(\pi,0)$.
The former boundary condition gives $(M_{\rm left}^z, M_{\rm right}^z)=(1/2,1/2)$ and the latter gives $(M_{\rm left}^z, M_{\rm right}^z)=(-1/2,-1/2)$.

We derived the symmetry-protected topological edge state in analogy with the edge magnetization of the spin-$1/2$ chain with the staggered field.
Both originate from the locking of the nonlocal field $\phi$ in the bulk.
The locking position $\bar\phi$ allows us to distinguish the topological Haldane phase from the trivial one.

To stand the nonzero edge magnetization, we need to lift the ground-state degeneracy.
An infinitesimal staggered magnetic field completely lifts the ground-state degeneracy.
The staggered field adds to the sine-Gordon Hamiltonian \eqref{H_SG_cos} the following interaction,
\begin{align}
    &h_s \sum_{j=1}^L (-1)^j \sum_{n=1,2}  S_{j,n}^z
    \notag \\
    &\approx a_1 h_s\int dx \, \bigl[\sin(2\phi_1) + \sin(2\phi_2) \bigr]
    \notag \\
    &= 2a_1h_s \int_0^L dx \sin(\phi_1+\phi_2) 
    \cos(\phi_1-\phi_2).
\end{align}
Note that the ferromagnaetic rung interaction also locks the antisymmetric mode, $\phi_1-\phi_2$, to $\phi_1-\phi_2=0\mod \pi$.
The locking of $\phi_1-\phi_2$ simplifies the staggered-field interaction at low energies:
\begin{align}
    h_s \sum_{j=1}^L (-1)^j \sum_{n=1,2} S_{j,n}^z
    &\propto h_s \int_0^L dx \, \sin \phi,
\end{align}
which makes the two locking positions $\bar\phi=\pi/2 \mod \pi$ and $\bar\phi=-\pi/2\mod \pi$ nonequivalent and lift the degeneracy of the spin-1/2 edge states.
In other words, $\mathcal P=-1/2$ and $\mathcal P=1/2$ are nonequivalent under the staggered field.
Therefore, while the quantized edge magnetization and the topological edge state have the same topological origin, the former becomes observable only after the edge-state degeneracy is lifted.

We comment that there is an alternative derivation of the edge magnetization in spin-1/2 ladders.
If $\cos(2\phi)$ has a scaling dimension $1$,  the sine-Gordon theory \eqref{H_SG_cos} can be refermionized and turned into a Majorana fermion theory~\cite{shelton_ladder}.
For $J_\perp<0$, these Majorana fermions are accompanied by zero-energy modes that give the edge magnetization $M_{\rm left}^z=\pm 1/2$ and $M_{\rm right}=\pm 1/2$~\cite{lecheminant_majorana_2002, orignac_ladder_2003, robinson_majorana_zero_2019}.

\subsection{Generalization to spin-$S$ chains}

\begin{figure}[t!]
    \centering
    \includegraphics[bb = 0 0 900 400, width=0.8\linewidth]{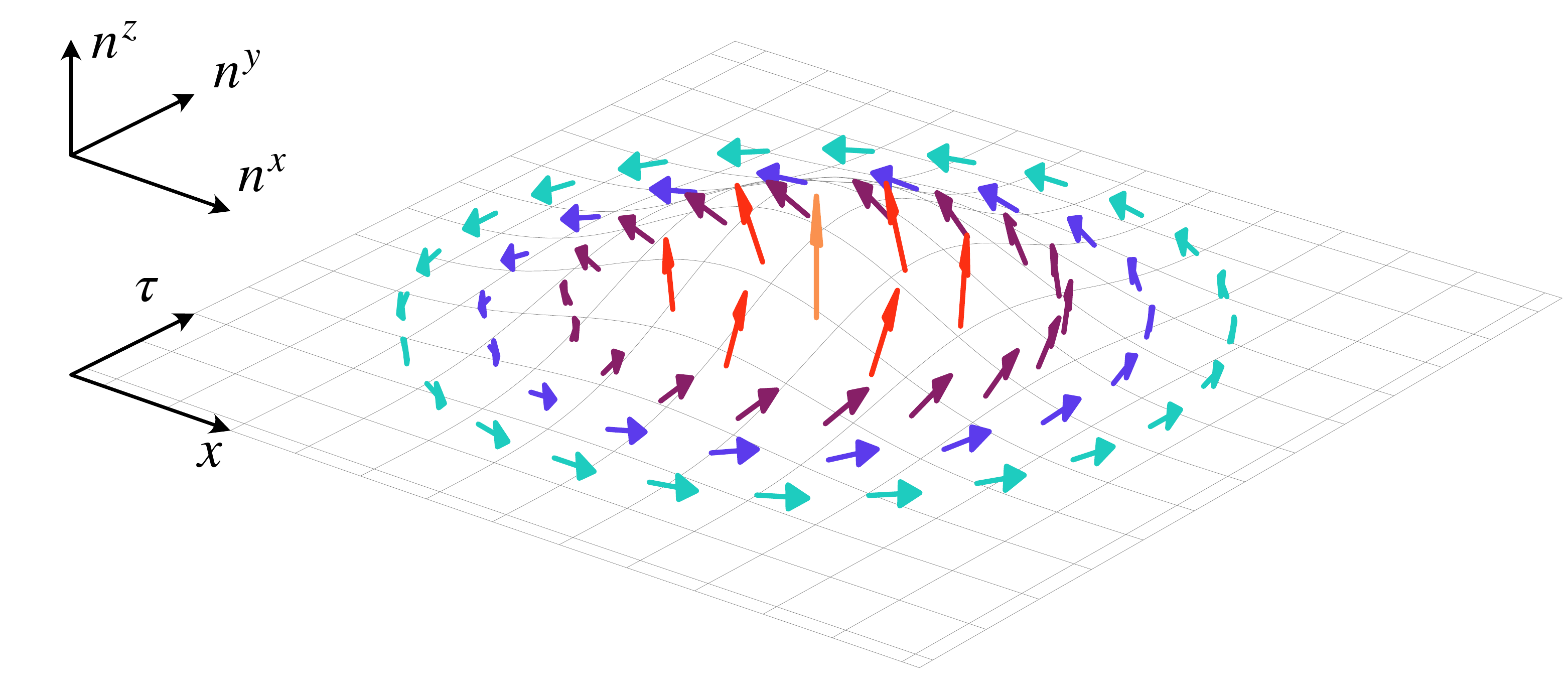}
    \caption{Meron configuration of staggered moment, $\bm n(x_j)=(-1)^j\bm S_j$, with skyrmion number $+1/2$.
    }
    \label{fig:meron}
\end{figure}

Our project is further continued to the general spin-$S$ HAFM chains.
There are two options to field theoretically discuss the spin-$S$ HAFM chains.
One is to use $2S$-leg spin-$1/2$ ladders~\cite{schulz_spin-S,fuji_spt_eft}, and the other is to use an O(3) nonlinear sigma model (NL$\sigma$M)~\cite{Affleck_review_1989,sachdev_textbook,auerbach_textbook}.
The latter is equivalent to the sine-Gordon theory thanks to a duality~\cite{affleck_meron} (Appendix~\ref{app:m=0}).
These dual theories are bridged by a topological excitation called a meron (Fig.~\ref{fig:meron})~\cite{gross_meron,affleck_meron}, which is a half of a skyrmion living in the two-dimensional Euclidean space-(imaginary) time.
The dual transformation from the O(3) NL$\sigma$M to the sine-Gordon theory is done in analogy with that for the two-dimensional XY model~\cite{KT_1973,Kosterlitz_1974}.

Here, we summarize the results obtained from the O(3) NL$\sigma$M approach.
We give technical details in Appendix.~\ref{app:m=0}.
The $S\in\mathbb{Z}+1/2$ cases fall into the $S=1/2$ case \eqref{H_SG}, where the staggered magnetic field induces the edge magnetization $\mathcal{P}=\pm1/4$.
With $h_s=0$, the half-odd-spin-$S$ HAFM chain has either the gapless ground state or doubly degenerate gapped ground state.
The sine-Gordon theory as the dual theory of the O(3) NL$\sigma$M explains this impossibility of the unique and gapped ground state under the translation and spin-rotation symmetries.
This argument is consistent with the LSM theorem on the spin-$S$ HAFM chain~\cite{lsm}.
The staggered magnetic field violates the one-site translation symmetry and makes the ground state unique and gapped.

The $S\in\mathbb{Z}$ cases are described by the sine-Gordon theory \eqref{H_SG_cos} with $g_c\propto -\cos(\pi S)$.
The edge magnetization $\mathcal{P}=\pm 1/2\mod1$ emerges only when $S\in2\mathbb{Z}+1$ [Eq.~\eqref{S_dual_m=0_2phi}].
By contrast, $\mathcal{P}=0\mod1$ follows from $\bar\phi=0\mod\pi$ for $S\in2\mathbb{Z}$.
This even-odd feature of the integer-spin-$S$ chain is consistent with the topological triviality (nontriviality) of the even-$S$ (odd-$S$) Haldane phase~\cite{pollmann_haldane_2012,tonegawa_spin-2}.

\subsection{Symmetry protection of quantization}

The quantization of the edge magnetization is protected by symmetries.
Typical interactions that ruin the quantization are U(1)-breaking interactions such as $J_x\sum_{j=1}^LS_j^xS_{j+1}^x$.
We can express this interaction in terms of the sine-Gordon theory as $\propto J_x\int_0^Ldx\,\cos(2\theta)$, where $\theta$ is a canonical conjugate of $\phi$~\cite{giamarchi_book} that satisfies a commutation relation,
\begin{align}
    [\phi(x), \theta(y)] = i\pi \Theta_{\rm step}(x-y),
    \label{comm_phi_theta}
\end{align}
where $\Theta_{\rm step}(z)$ is the Heaviside step function,
\begin{align}
    \Theta_{\rm step}(z)
    =\left\{
    \begin{array}{ccc}
        1 & & (z>0)  \\[+4pt]
        \frac 12 & & (z=0) \\[+4pt]
        0 & & (z<0)
    \end{array}.
    \right.
\end{align}
Equation~\eqref{comm_phi_theta} indicates that $\phi$ and $\theta$ are not locked at the same time.

Our ``semiclassical bosonization'' formula of the spin derived from the spin chain through the O(3) NL$\sigma$M (Appendix~\ref{app:bosonization_m=0}) leads to $(-1)^jS_j^+\approx e^{i\theta}$ for the spin-$S$ chain just like the conventional one~\cite{giamarchi_book}.
The global U(1) spin-rotation symmetry excludes $\cos(n\theta)$ and $\sin(n\theta)$ with $n\in\mathbb{Z}$ from the Hamiltonian.

Fixing the locking position $\bar\phi$ is also necessary to protect the quantization.
If the effective Hamiltonian \eqref{H_SG} for the spin-$1/2$ chain admits an interaction $g_c\cos(2\phi)$, 
the Hamiltonian is modified to
\begin{align}
	\mathcal H_s = \int_0^Ldx \biggl[ \frac v{2\pi K}  (\partial_\mu\phi)^2 + \sqrt{{g_s}^2+{g_c}^2} \sin(2\phi + A) \biggr],
\end{align}
with $A=\tan^{-1}(g_c/g_s)$.
The locking position gets shifted in accordance with the shifted potential $\sqrt{{g_s}^2+{g_c}^2}\sin(2\phi+A)$.
The bulk polarization $\mathcal{P}$ gradually changes with $A$~\cite{shindou_pump}.
Hence, we must forbid $\cos(2\phi)$ from entering into the spin-$1/2$ Hamiltonian \eqref{H_SG} to keep $\mathcal P$ quantized.
Likewise, we must forbid $\sin(2\phi)$ from the spin-1 Hamiltonian \eqref{H_SG_cos}.

An inversion symmetry fixes the locking position of $\phi$.
For the spin-1/2 chain, the site-centered inversion $\mathcal{I}_s:\,\bm{S}_j\to\bm{S}_{L-j}$ acts on $\phi$ as $\mathcal{I}_s:\,\phi(x)\to-\phi(L-x)+\pi/2$.
This operation on $\phi$ is deduced from behaviors of the staggered magnetization $(-1)^j S_{j}^z\sim \sin(2\phi)$ and the dimerization $(-1)^j\bm S_j \cdot \bm S_{j+1}\sim \cos(2\phi)$.
The site-centered inversion keeps the former invariant but changes the sign of the latter.
The $\mathcal{I}_s$ symmetry thus excludes $\cos(2\phi)$ from the spin-1/2 chain \eqref{H_SG}.
For the spin-1 chain, $\mathcal{I}_s$ instead acts on $\phi=\phi_1+\phi_2$ so that $\mathcal{I}_s:\,\phi(x)\to-\phi(L-x)+\pi$ because both $\phi_1$ and $\phi_2$ admit the $\pi/2$ shift under $\mathcal{I}_s$.
The $\mathcal{I}_s$ symmetry thus excludes $\sin(2\phi)$ from the spin-1 chain \eqref{H_SG_cos}.
Based on our semiclassical bosonization formulas for the spin-$S$ HAFM chains,
we reach the same conclusion that the U(1) spin-rotation and the $\mathcal I_s$ inversion symmetries protect the quantization of the edge magnetizations in the spin-$S$ HAFM chains.

The symmetry-protected edge magnetization gives an interesting characterization of quantum phases in one-dimensional quantum spin systems.
Here, we point out that the edge-magnetization-based characterization of quantum phases is related to the concept of symmetry-protected trivial (SPt) phases~\cite{fuji_sptrivial}.
Though the precise relation between these two concepts is not clear yet, they are related indeed.
According to Ref.~\cite{fuji_sptrivial}, the induced N\'eel state $\ket{N} = \bigotimes_j \ket{S_j^z=(-1)^j}$ in a spin-$1$ chain is an SPt phase but the large-$D$ state $\ket{D}=\bigotimes_j\ket{S_j^z=0}$ is not.
The former has $\mathcal P= 1/2 \mod 1$ and the latter has $\mathcal P=0 \mod 1$, since $\braket{N|U|N} = -1$ and $\braket{D|U|D}=1$.
Note that the spin-$1/2$ chain \eqref{H_chain} has $\ket{N}$ as its ground state in the $h_s\to+\infty$ limit.
We will summarize affinities and differences of our characterization of quantum phases with the odd-spn Haldane phase and the SPt phase later in Table.~\ref{tab}.

\section{Edge magnetization on magnetization plateau}
\label{sec:plateau}

The edge-magnetization-based characterization of quantum phases also works when the net magnetization is nonzero, $\sum_{\bm r}\braket{S_{\bm r}^z}\not=0$.
Generally, the uniform magnetic field favors spatially nonuniform $\phi$ accompanied by gapless spin excitations [see Eq.~\eqref{MzNz}].
The uniform magnetic field $h_u$ indeed reduces the Haldane gap of the spin-1 chain and induces the quantum phase transition into a TL-liquid phase~\cite{giamarchi_book,schulz_c_ic,chitra-giamarchi}.
However, the uniform magnetic field can also induce a quantum phase transition from the TL liquid phase into a spin gap phase.
An increase of the uniform magnetic field can yield a spin gap state with a commensurate magnetization density, called a magnetization plateau~\cite{oya}.
Magnetization plateaus are supposed to appear when $N_0(S-m)$ is a rational number~\cite{oya}, where $N_0$ is the number of spins per unit cell and $m$ is the magnetization per site.

\subsection{Spin-1/2 ladder}

\begin{figure}[t!]
    \centering
    \includegraphics[bb = 0 0 900 500,width=\linewidth]{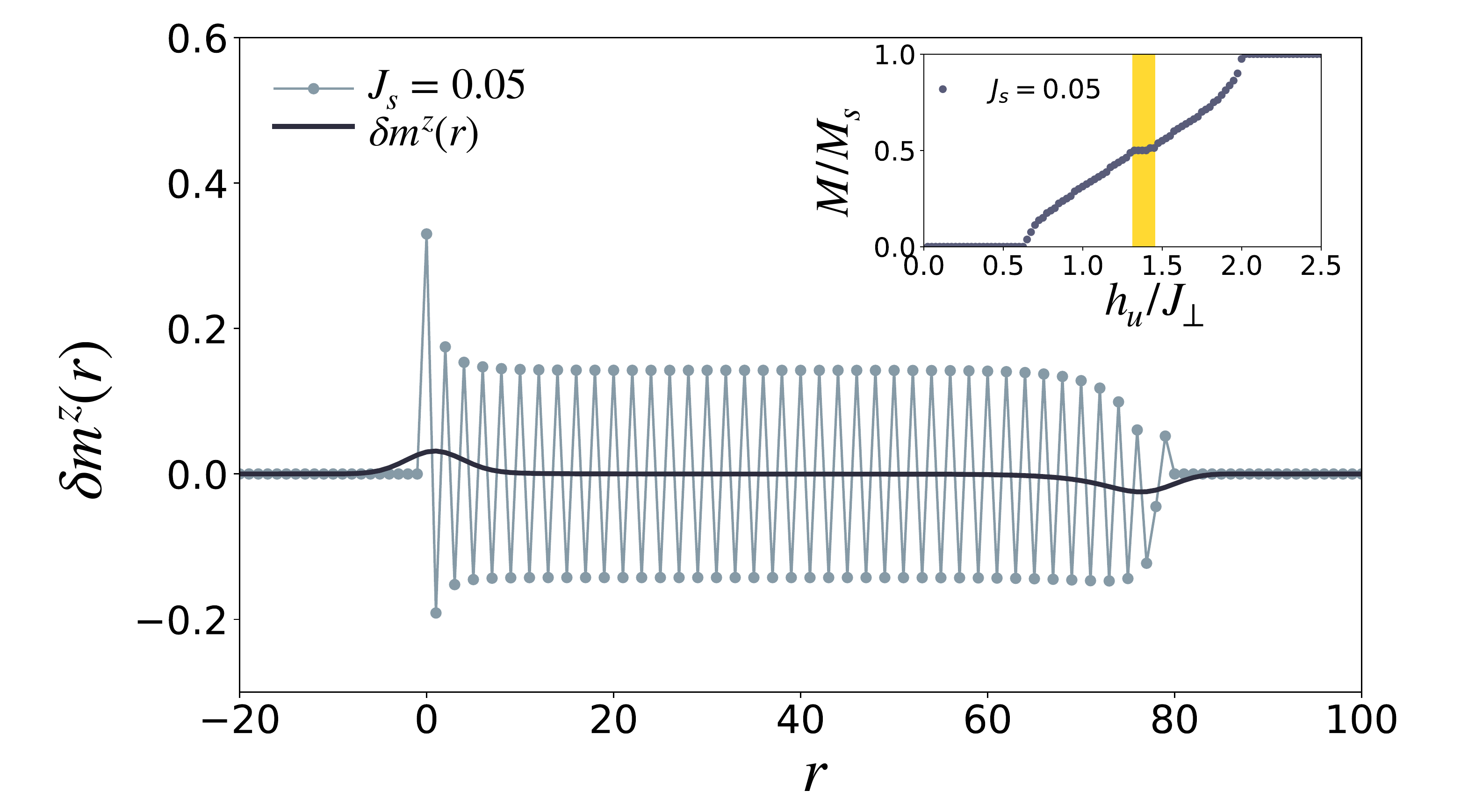}
    \caption{Site $r$ dependence of the ``charge'' $C_r$ of Eq.~\eqref{U_def_plateau} with $N=2$ (filled balls) and $\delta m^z(r)$. 
    The edge magnetizations on the left and right edges are quantized as $\pm 0.25000$.
    We used parameters $J_\perp=1$, $J_\parallel=0.5$, $J_s=0.05$, $h_u=1.4$, and $2L=160$.
    (inset) Magnetization curve. The shaded area highlights the $1/2$ magnetization plateau.
    }
    \label{fig:ladder_edge}
\end{figure}

Figure~\ref{fig:ladder_edge} shows numerical results about a spin-$1/2$ ladder that exhibits a magnetization plateau.
Its Hamiltonian is given by,
\begin{align}
    \mathcal{H}_{\mathrm{ladder}}
    &=J_\parallel\sum_{j=1}^{L-1}\sum_{n=1,2}\bm{S}_{j,n}\cdot\bm{S}_{j+1,n}-h_u\sum_{j=1}^L\sum_{n=1,2}S_{j,n}^z
    \notag \\
    &\quad +\sum_{j=1}^L[J_\perp+(-1)^jJ_s]\bm{S}_{j,1}\cdot\bm{S}_{j,2},
    \label{H_ladder}
\end{align}
with $J_\perp \gg  J_\parallel>0$, $J_\perp \gg J_s >0$, and $h_s>0$.
The first term denotes the intra-chain HAFM interaction, the second term is the uniform Zeeman field, and the others are the uniform ($J_\perp$) and the staggered ($J_s$) rung interactions.
All the numerical data in this paper are obtained from density-matrix renormalization group calculations with the OBC~\cite{itensor}, where the bond dimension $100-1000$  and the truncation error $10^{-8}$ at largest are kept.
While the spin ladder with $J_s=0$ does not exhibit the magnetization plateau, the staggered rung interaction, $J_s\sum_{j=1}^L(-1)^j\bm{S}_{j,1}\cdot\bm{S}_{j,2}$, induces the magnetization plateau with $M/M_s=1/2$, where $M_s$ is the saturated value of the magnetization (the inset of Fig.~\ref{fig:ladder_edge})~\cite{tonegawa_ladder_plateau}.

\subsection{Pseudospin picture}

The ground state of the model \eqref{H_ladder} on the $1/2$ plateau has an intuitive pseudospin picture~\cite{giamarchi_book}.
For $J_\parallel=0$, the spin ladder \eqref{H_ladder} is fractured into a set of isolated antiferromagnetic dimers.
Each dimer is described by a two-spin model 
\begin{align}
    \mathcal H_{\text{dimer}}=J_\perp\bm{S}_1\cdot\bm{S}_2-h_u(S_1^z+S_2^z).
    \label{H_2spin}
\end{align}
If $h_u=0$, the ground state of the model \eqref{H_2spin} is the singlet state, $\ket{s}=(\ket{\frac{1}{2},\,-\frac{1}{2}}-\ket{-\frac{1}{2},\,\frac{1}{2}})/\sqrt{2}$, 
where $\ket{S_1^z,S_2^z}$ is the simultaneous eigenstate of $S_n^z$ for $n=1,2$.
The dimer has triply degenerate spin-1 excited states, $\ket{t_\sigma}$ with $\sigma=1,0,-1$.
The uniform magnetic field $h_u$ lifts the triple degeneracy  and eventually makes $\ket{s}$ and $\ket{t_1}=\ket{\frac{1}{2},\,\frac{1}{2}}$ nearly degenerate, where we may discard the other high-energy excited states and regard $\ket{s}$ and $\ket{t_1}$ as the ``down'' and ''up'' states of an $S=1/2$ pseudospin $\bm{T}_j$~\cite{giamarchi_ladder_pseudospin,bouillot_bpcb}.

We can regard the spin ladder \eqref{H_ladder} as a weakly coupled dimers for $J_\parallel/J_\perp \ll 1$ and, accordingly, as a pseudospin-$1/2$ chain.
The perturbative expansion about $J_\parallel/J_\perp$ maps the spin-$1/2$ ladder \eqref{H_ladder} into the single pseudospin-$1/2$ XXZ chain with a staggered magnetic field, 
\begin{align}
    \mathcal{H}_{\mathrm{ladder}}
    &\approx
    J_\parallel\sum_{j=1}^L(T_j^xT_{j+1}^x+T_j^yT_{j+1}^y+\tfrac{1}{2}T_j^zT_{j+1}^z)
    \notag \\
    &\qquad -\sum_{j=1}^L(h_{\mathrm{eff}}-(-1)^jJ_s)T_j^z,
\end{align}
with $h_{\rm eff} = h_u - J_\perp-\frac{J_\parallel}{2}$.
The effective uniform magnetic field vanishes, $h_{\mathrm{eff}}=0$, for $M/M_s=1/2$ if $J_s=0$~\cite{bouillot_bpcb}.
Nonzero $J_s$ opens the spin gap around $h_{\rm eff}\approx0$.
In other words, the ground state is on the $1/2$ magnetization plateau around $h_{\rm eff}\approx 0$ for $J_s\not=0$.
Similarly to the authentic spin-1/2 chain, the edge magnetization $\pm1/4$ will appear in the pseudospin-$1/2$ chain (Fig.~\ref{fig:ladder_edge}).

\subsection{General definition of edge magnetization on magnetization plateaus}\label{sec:gen_mp}

Let us formulate the edge magnetization on the magnetization plateau based on a nonlocal operator $U$.
A naive definition of the operator $U$ in the spin ladder will be
\begin{align}
    U_{\rm naive} = \exp \biggl( i\frac{2\pi}{L}\sum_{j=1}^Lj \sum_{n=1}^{N_0} S_{j,n}^z\biggr),
    \label{U_plateau_naive}
\end{align}
where $S_{j,n}^z$ is the $z$ component of a spin $\bm S_{j,n}$ in the unit cell located at the position $j$.
The unit cell contains $N_0=4$ spins for $J_s\not=0$.
The operator $U_{\rm naive}$ of Eq.~\eqref{U_plateau_naive} seems ambiguous on the magnetization plateau because, if we regard $S_j^z$ as a charge, the charge neutrality condition \eqref{neutral_Sz} is violated on the magnetization plateau.
Suppose that the magnetization per site $m$ is fractional, namely $m=p/q$ with coprime positive integers $p$ and $q$.
The naive operator \eqref{U_plateau_naive} gives $\braket{U_{\rm naive}}= \exp(2\pi mi (L+1))$.
Since $m$ is fractional, $U_{\rm naive}$ formally leads to $\mathcal P\not=0$.
However, this $\mathcal P\not=0$ cannot be deemed the edge magnetization because it originates from the bulk uniform magnetization.
To make the edge magnetization well defined on the nonzero magnetization plateau, we need to define an appropriate charge that satisfies a charge neutrality condition.
In general one-dimensional quantum spin systems on magnetization plateaus, we can adopt
\begin{align}
    U = \exp\biggl(i\frac{2\pi}{L}\sum_{j=1}^L jC_j\biggr), \quad C_j = \sum_{n=1}^N (S_{j,n}^z-m),
    \label{U_def_plateau}
\end{align}
where $C_j$ is the charge density for $m\not=0$.
The naive choice of $N$ will be $N=N_0$, the number of spins per unit cell.
However, $N$ can be smaller than $N_0$.
Recall that we took $N=1$ for the spin-$1/2$ chain \eqref{H_chain} despite $N_0=2$.
We take $N=2$ for the spin ladder \eqref{H_ladder} even for $J_s\not=0$, where $N_0=4$.
The magnetization plateau satisfies the charge neutrality,
\begin{align}
    \sum_{j=1}^L\braket{C_j}=0.
    \label{neutrality_C}
\end{align}
We modify the definition of the edge magnetizations in accordance with the definition of the charge $C_j$.
\begin{align}
    M_{\mathrm{left}}^z &= \int_{-\infty}^{\frac{L+1}2} dr \,\delta m^z(r), \quad M_{\mathrm{right}}^z = \int_{\frac{L+1}2}^\infty dr \,\delta m^z(r),
    \label{Ml_Mr_plateau} \\
    \delta m^z(r) &= \sum_{r'=1}^L g(r-r') \braket{C_{r'}}.
    \label{mz_plateau}
\end{align}

For the spin ladder \eqref{H_ladder} on the $1/2$ plateau, Eq.~\eqref{U_def_plateau} gives 
\begin{align}
    U=\exp\biggl( i\frac{2\pi}{L}\sum_{j=1}^L j T_j^z \biggr),
\end{align}
dealing with the pseudospin-$1/2$ chain in analogy with the authentic spin-$1/2$ chain.
Figure~\ref{fig:ladder_edge} shows the spatial distribution of $\braket{C_r}$ and $\delta m^z(r)$ for the spin-$1/2$ ladder \eqref{H_ladder} on the $1/2$ magnetization plateau .
The edge magnetizations \eqref{Ml_Mr_plateau} exhibit the excellent quantization $M_{\mathrm{left}}^z=-M_{\mathrm{right}}^z=0.25000$ despite the small $J_s\ll J_\perp$.
We used the Gaussian $g(r)$ with $\lambda=3$.
The Lorentzian $g(r)=(\varepsilon/\pi)/(r^2+\varepsilon^2)$ also exhibits the edge magnetization but its quantization is less accurate.
We also confirmed that the quantization accuracy is insensitive to the value of $\lambda$.

\begin{figure}
    \centering
    \includegraphics[bb = 0 0 1050 1100, width=\linewidth]{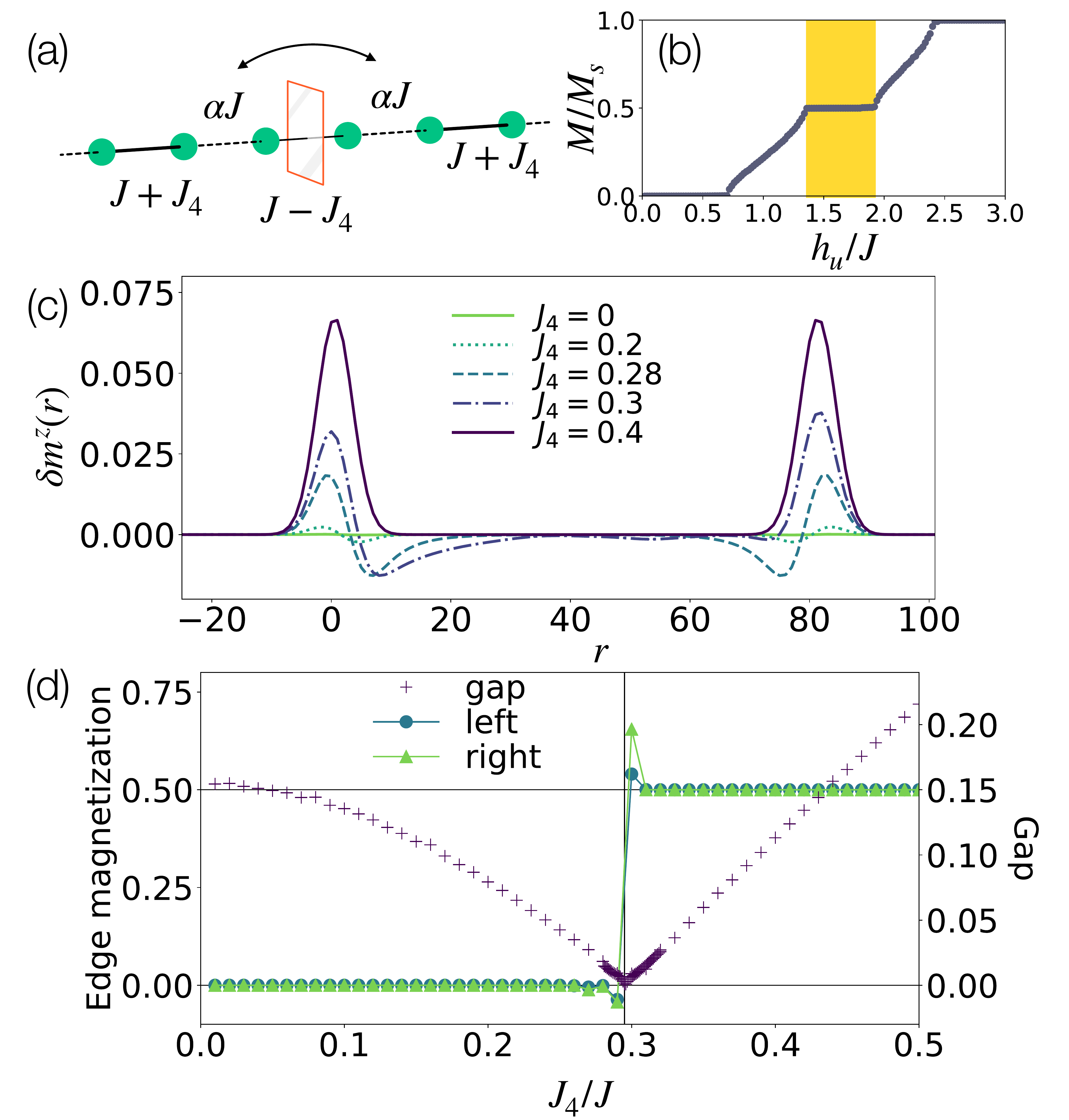}
    \caption{(a) Spin-1 chain \eqref{H_4} with bond-inversion center. (b) Magnetization curve for $J_4=0$. (c) Site $r$ dependence of $\delta m^z(r)$
    (d) $J_4$ dependence of lowest-energy excitation gap ($+$ markers) and edge magnetizations~\eqref{mz_plateau} (circles and triangles).
    The edge magnetizations are quantized as $M_{\mathrm{left}}^z=M_{\mathrm{right}}^z=0.499999999$ for $J_4>J_{4c}\approx0.295$.
    We used parameters $J=1$, $\alpha=0.2$, $h_u=1.5$.
    The system size is $L=162$ for (b,c) and $242$ for (d).
    }
    \label{fig:bondalt_edge}
\end{figure}

\section{Charge jump at quantum critical point}
\label{sec:J4}

The spin-$1/2$ ladder \eqref{H_ladder} is the prototypical model that exhibits the quantized edge magnetizations on the nonzero magnetization plateau.
In what follows, we consider a thought-provoking model whose edge magnetization jumps from zero to a nonzero value at a quantum critical point on the plateau.

\subsection{Four-site periodic spin-1 chain and its pseudospin picture}

The model is a spin-1 HAFM chain with a four-site periodic structure [Fig.~\ref{fig:bondalt_edge}~(a)],
\begin{align}
    \mathcal{H}_{4}
    &=J\sum_{j=1}^{\frac{L-2}2} (\bm S_{2j-1}\cdot \bm S_{2j}+\alpha\bm S_{2j}\cdot\bm S_{2j+1})
    \notag \\
    &\qquad + J_4 \sum_{j=1}^{\frac L2}(-1)^j \bm S_{2j-1} \cdot \bm S_{2j} - h_u \sum_j S_j^z,
    \label{H_4}
\end{align}
where $\bm S_j$ is the spin-1 operator at the $j$th site.
$0\le \alpha \le 1$ and $0\le J_4 < J$.
The parameter $\alpha$ denotes the bond alternation for $\alpha\not=1$.
The $J_4$ interaction is the four-site periodic exchange interaction.
We can label $\bm{S}_{2j-1}=\bm{S}_{r,1}$ and $\bm{S}_{2j}=\bm{S}_{r,2}$ for $r=1,2, \cdots, L/2$.
Here, we define the charge \eqref{U_def_plateau} as
\begin{align}
    C_r = \sum_{n=1}^2 (S_{r,n}^z-m).
\end{align}

The bond-alternating chain \eqref{H_4} shows magnetization plateaus [Fig.~\ref{fig:bondalt_edge}~(b)].
The $1/2$ plateau for $J_4=0$ was experimentally observed~\cite{narumi_ntenp2}.
We can see this $1/2$ plateau as a forced ferromagnetic phase of another $S=1/2$ pseudospin, $\tilde{\bm{T}}_j$.
For $\alpha=0$, the model \eqref{H_4} is reduced to isolated spin-1 dimers.
Each dimer is described by the Hamiltonian \eqref{H_2spin}.
This time, the spin quantum number of $\bm S_1$ and $\bm S_2$ are $S=1$.
For $h_u=0$, each dimer has the spin-0 singlet ground state, $\ket{s}=(\ket{1,-1}+\ket{-1,1}-\ket{0,0})/\sqrt{3}$.
The magnetic field reduces the excitaton energy of a spin-1 state $\ket{t_1}=(\ket{1,0}-\ket{0,1})/\sqrt{2}$ and makes $\ket{s}$ and $\ket{t_1}$ degenerate at $M/M_s=1/4$.
Regarding $\ket{s}$ and $\ket{t_1}$ as the ``down'' and ''up'' state of the $S=1/2$ pseudospin $\tilde{\bm{T}}_j$, we can rewrite the Hamiltonian \eqref{H_4} as
\begin{align}
    \mathcal{H}_4
    &\approx \frac{J\alpha}{3}\sum_j(\tilde{T}_j^x\tilde{T}_{j+1}^x+\tilde{T}_j^y\tilde{T}_{j+1}^y+\tfrac{3}{4}\tilde{T}_j^z\tilde{T}_{j+1}^z)
    \notag \\
    &\qquad -\sum_{j}(h'_{\mathrm{eff}}-(-1)^jJ_4)\tilde{T}_j^z
\end{align}
within the first-order approximation about $\alpha$~\cite{okamoto_bip-teno}.
The $J_4$ interaction turns into the staggered magnetic field, which induces the $1/4$ magnetization plateau around $h'_{\mathrm{eff}}=0$.
As $h'_{\mathrm{eff}}$ is increased from zero, the pseudospin chain eventually reaches the forced ferromagnetic phase of the pseudospin, that is, the $1/2$ magnetization plateau with $\braket{S_{2j-1}^z}=\braket{S_{2j}^z}=(1+2\braket{\tilde{T}_j^z})/4=1/2$.
The $1/2$ magnetization plateau thus exists even for $J_4=0$~\cite{totsuka_bondalt_plateau}.

To understand the magnetization process for $M/M_s>1/2$, we need to go beyond this pseudospin approximation.
For $M/M_s>1/2$, a spin-2 state $\ket{q_2}=\ket{1,1}$ of the spin-1 dimer enters into the ground state.
The inter-dimer interactions make $\ket{q_2}$ dispersive.
The magnetization process for $1/2<M/M_s\le1$ corresponds to a process that increases a population of $\ket{q_2}$ in the ground state.

\subsection{Quantum phase transition on 1/2 magnetization plateau}

The effective staggered magnetic field $J_4(-1)^j\tilde{T}_j^z$ would seem to drag the ground state away from the $1/2$ plateau since the forced N\'eel state $\bigotimes_{j=1}^{L/2}\ket{\tilde{T}_j^z=(-1)^{j+1}/2}$ in accordance with the effective staggered-field interaction has $M/M_s=1/4$.
However, the $J_4$ interaction induces a quantum phase transition without dragging the ground state away from the $1/2$ plateau.
Figure~\ref{fig:bondalt_edge}~(c) shows the $J_4$ dependence of $M_{\mathrm{left}}^z$ and $M_{\mathrm{right}}^z$.
We chose $L=162$ to keep the bond-centered inversion symmetry, $\mathcal{I}_b:\,\bm{S}_j\to\bm{S}_{L+1-j}$ under the OBC.
The model \eqref{H_4} is $\mathcal{I}_b$-invariant only when $L=2\mod4$~\footnote{The chain length $L$ must be even to guarantee the charge neutrality, as we mentioned in Sec.~\ref{sec:zero_field_1/2}. The model \eqref{H_4} has the bond-centered inversion symmetry, $\mathcal I_b:\bm S_j \to \bm S_{L+1-j}$ for $L=2\mod 4$ but does not have the site-centered inversion symmetry, $\mathcal I_s:\bm S_j \to \bm S_{L-j}$. By contrast, the same model has the $\mathcal I_s$ symmetry but does not have the $\mathcal I_b$ symmetry for $L=0 \mod 4$.}.
The U(1) spin-rotation and $\mathcal I_b$ symmetries protect the quantization of the edge magnetization, as we see later.
The $\mathcal I_s$ cannot protect the quantization of the edge magnetization.
The spatial dependence $\delta m^z(r)$ gradually changes as $J_4$ is increased [Fig.~\ref{fig:bondalt_edge}~(c)].
Nevertheless, the edge magnetizations shows a jump from zero to $1/2$ at $J_4=J_{4c}\approx 0.295J$ [Fig.~\ref{fig:bondalt_edge}~(d)] since the above symmetries forbid the continuous change [Eq.~\eqref{U_Ib}].

An effective field theory gives a straightforward way to understand the charge jump.
It was previously shown that the effective field theory on the magnetization plateau becomes the sine-Gordon theory for $S-m\in\mathbb{Z}$~\cite{tanaka_plateau_eft,takayoshi_plateau_eft}.
This sine-Gordon theory itself is inapplicable to the current situation of our interest with $S-m=1/2$.
However, we can modify the argument of Refs.~\cite{tanaka_plateau_eft,takayoshi_plateau_eft} to fit into our situation (Appendix~\ref{app:S-m=1/2}).
We can resolve the issue of the fractional $S-m$ by properly counting topological sectors.
The previously considered case with $S-m\in \mathbb Z$ involves the single topological sector.
Our case with $S-m\in\mathbb{Z}+1/2$ involves two topological sectors.
This topological difference governs the behavior of the effective field theory.

Let us first discuss the effective field theory in the bulk by imposing the periodic boundary condition, $\bm S_{j+L}=\bm S_j$, on the spin chain \eqref{H_4}.
When $\alpha-1=J_4=0$, the spin chain \eqref{H_4} is the uniform spin-1 HAFM chain.
When $S-m=1/2$, the spin-1 chain is described by an effective field theory with the following Hamiltonian [see Eq.~\eqref{S_dual_S-m=1/2}]:
\begin{align}
    \mathcal H_{\rm eff} &\approx \frac v{2\pi K} \int dx \, (\partial_\mu \phi)^2
    \notag \\
    &\qquad - 2\zeta^2 \cos[2\pi (S-m)] \int dx \, \cos(4\phi),
    \label{H_4_eff_uniform}
\end{align}
where $\zeta$ is a fugacity of a vortex~\cite{tanaka_plateau_eft,takayoshi_plateau_eft}.
When the $\cos(4\phi)$ interaction is relevant, the spin-1 chain shows the magnetization plateau by spontaneously breaking the one-site translation symmetry, $T_1:\bm S_{j} \to \bm S_{j+1}$.
When the $\cos(4\phi)$ interaction is irrelevant, the spin-1 chain has the gapless TL-liquid ground state for $S-m=1/2$.
While it is hard to judge which is the case only from the effective field theory, we know numerically that the gapless scenario is the case~\cite{takahashi_haldane_mag}.
Hence, we may drop the $\cos(4\phi)$ term from the Hamiltonian \eqref{H_4_eff_uniform}.

The $T_1$ symmetry is translated into Eq.~\eqref{T1_S-m=1/2_phi}, namely,
\begin{align}
    \phi(x) \to \phi(x) + \frac\pi 2, \\
    \theta(x) \to \theta(x) +\pi,
\end{align}
in the field theory language, which excludes $\cos(2\phi)$ and $\sin(2\phi)$ from the Hamiltonian.
The absence of $\cos(2\phi)$ and $\sin(2\phi)$ is due to the destructive interference between the two topological sectors.
The bond alternation $\alpha\not=1$ and the $J_4$ interaction break the $T_1$ symmetry and unbalance the interference between the topological sectors.
The model \eqref{H_4} on the $1/2$ plateau is mapped to the sine-Gordon theory,
\begin{align}
    \mathcal{H}_4 &= \int_0^Ldx\biggl[\frac{v}{2\pi K} (\partial_\mu\phi)^2 +(g_2(J_4)-g_{2c})\cos(2\phi)
    \biggr],
    \label{H_SG_4}
\end{align}
with $g_2(J_4)\propto (J_4)^2/J$ and $g_{2c}\propto(1-\alpha)J$.
The $J_4$ interaction of Eq.~\eqref{H_4} is taken into account perturbatively.
The $J_4$ interaction gives rise to $\cos(2\phi)$ in a second-order perturbation process (Appendix~\ref{app:J4}) because the $J_4$ interaction has the four-site periodicity.
The cosine interaction $\cos(2\phi)$ is the most relevant interaction with the two-site periodicity in accordance with the antiferromagnetic fluctuations.
The second-order perturbation expansion is required to produce the two-site periodic interaction from the four-site one.
Note that $\sin(2\phi)$ is still forbidden because of the $\mathcal I_b$ symmetry, $\phi(x) \to -\phi(L-x) \mod 2\pi$ [Eq.~\eqref{Ib_J4_chain}].

Next, we impose the OBC on the spin chain \eqref{H_4}.
The $T_1$ symmetry is lost, but the $\mathcal I_b$ symmetry survives in the OBC.
The effective Hamiltonian \eqref{H_SG_4} thus holds with the OBC as well as the periodic one.
When $J_4=0$, the sine-Gordon theory \eqref{H_SG_4} with $g_{2c}>0$ locks $\bar\phi=0$ (i.e. $\mathcal{P}=0$).
$J_4\not=0$ makes $g_2(J_4)>0$.
An increase of $J_4>0$ drags the system into the quantum critical point $J_4=J_{4c}$, where $g_2(J_4)=g_{2c}$ holds.
For $J_4>J_{4c}$, the sine-Gordon theory \eqref{H_SG_4} with $g_2(J_4)-g_{2c}<0$ locks $\bar\phi=\pi/2$ (i.e. $\mathcal{P}=1/2$).

\subsection{Topological transition on magnetization plateau}

\begin{figure}[t!]
    \centering
    \includegraphics[bb = 0 0 900 900, width=\linewidth]{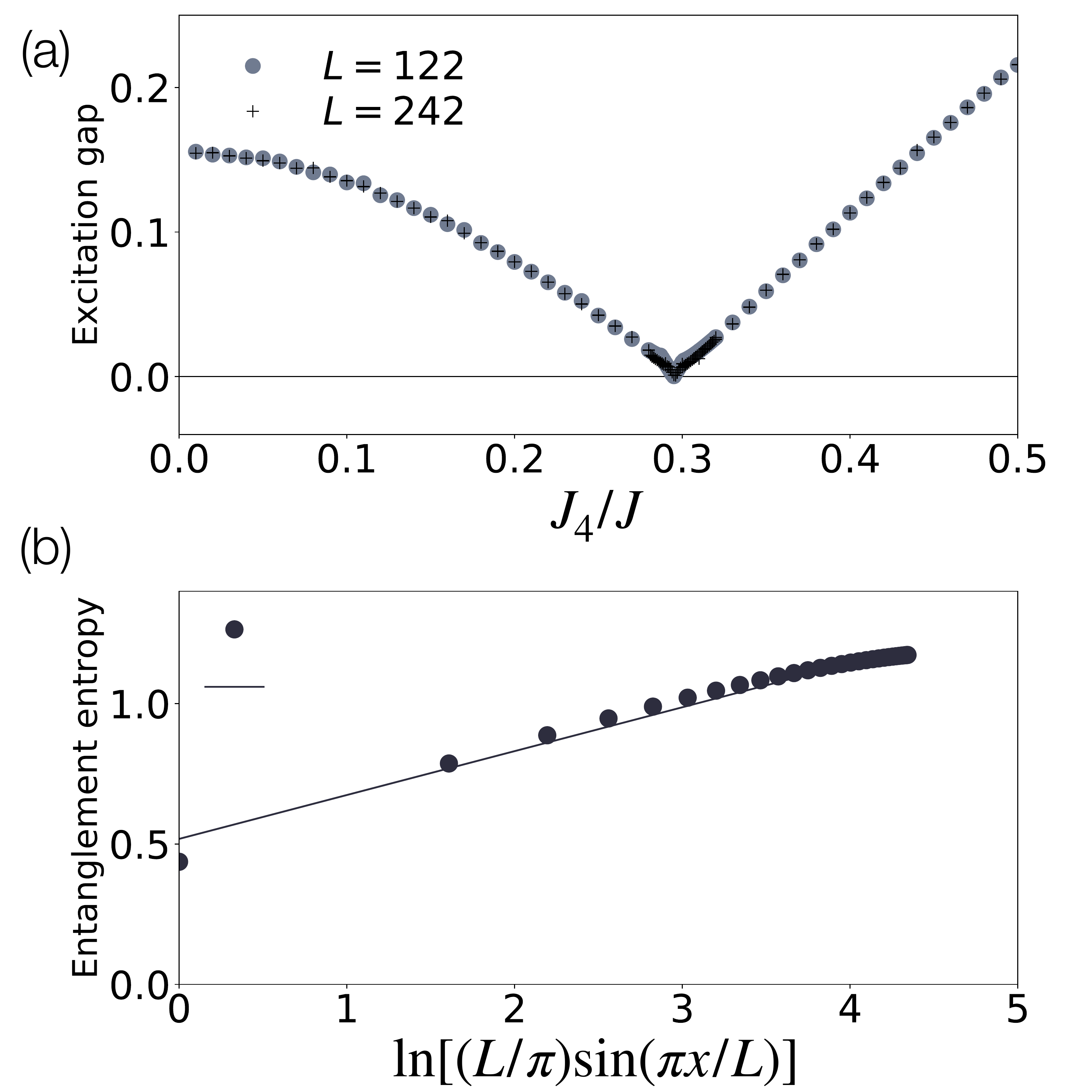}
    \caption{(a) $J_4$ dependence of lowest-energy excitation gap for $h_u/J=1.5$, $\alpha=0.2$, and $L=122$ (circles) and $L=242$ ($+$ markers).
    (b) Numerically calculated entanglement entropy for $h_u/J=1.5$, $\alpha=0.2$, and $L=242$ (circles) and the fitting function \eqref{S_EE} with $(a_s,c) = (0.52, 0.94)$. The central charge $c\approx 1$ is consistent with the sine-Gordon theory \eqref{H_SG_4}.
    }
    \label{fig:gap_ee}
\end{figure}

Numerical results show that the quantum phase transition at $J_4=J_{4c}$ is likely to be a quantum critical one with the gap closing [Fig.~\ref{fig:bondalt_edge}~(d), Fig.~\ref{fig:gap_ee}~(a)].
In the quantum critical regime, 
the entanglement entropy 
shows the logarithmic scaling~\cite{calabrese_ee} [Fig.~\ref{fig:gap_ee}~(b)],
\begin{align}
    \mathcal S_{\rm EE}(x) &= a_s + \frac c6 \ln \biggl[\frac{L}{\pi}\sin\biggl(\frac{\pi x}{L}\biggr)
    \biggr],
    \label{S_EE}
\end{align}
with a constant $a_s$ and the central charge $c$.
The central charge characterizes the conformal field theory that corresponds to the quantum critical point.
The numerical estimation $c\approx 0.94$ is consistent with $c=1$ of the effective field theory \eqref{H_SG_4} with $g_2(J_4)=g_{2c}$.
The gap is thus highly likely to be closed at the transition point $g_2(J_4)=g_{2c}$.
Even if the phase transition should be weakly first-order, our field theory \eqref{H_SG_4} still holds by taking into account a less relevant interaction, $\cos(4\phi)$, which is dropped in Eq.~\eqref{H_SG_4} (see Eq.~\eqref{H_4_phi}).
If the first-order transition scenario comes true, the $J_4$ interaction must affect the scaling dimension of the $\cos(4\phi)$ interaction to make it relevant.

Despite the ground state's quantum critical behavior, the edge magnetization shows the abrupt change, reflecting its topological nature.
The quantization of the edge magnetization is an exact property of the spin chain \eqref{H_4}.
The $\mathcal{I}_b$ symmetry of the Hamiltonian imposes that 
\begin{align}
    \braket{U} =\braket{\mathcal{I}_bU\mathcal{I}_b^{-1}}=\braket{U^\dag}.
\end{align}
It follows that $\im\ln\braket{U}=-\im\ln\braket{U}\mod2\pi$. Namely, the $\mathcal I_b$ imposes
\begin{align}
    \mathcal P &= 0, \> \text{ or } \> \frac 12 \mod 1.
\end{align}
This is consistent with the numerical results.
$\mathcal P=0$ holds for $J_4<J_{4c}$ and $\mathcal P=1/2$ for $J_4>J_{4c}$.

Our quantum field theory \eqref{H_SG_4} also supports the fact that the U(1) spin-rotation and the $\mathcal I_b$ symmetries protect the quantization of the edge magnetization.
According to Eqs.~\eqref{Ib_J4_chain}, the $\mathcal I_b$ symmetry forbids $\sin(n\phi)$ with $n \in \mathbb Z$ from entering into the effective Hamiltonian.
Besides, the U(1) spin-rotation symmetry forbids $\cos(n\theta)$ and $\sin(n\theta)$.
Therefore, the quantization of the edge magnetization of the spin-1 chain \eqref{H_4} on the $1/2$ plateau is protected by the U(1) spin-rotation and the $\mathcal I_b$ symmetry.

Precisely speaking, the ground state for $J_4>J_{4c}$ violates the charge neutrality condition \eqref{neutrality_C} because of
$M_{\mathrm{left}}^z=M_{\mathrm{right}}^z=1/2$ [Figs.~\ref{fig:bondalt_edge}~(c), (d)].
$\sum_{j=1}^L\braket{C_j}=1$ holds for $J_4>J_{4c}$ whereas $\sum_{j=1}^L\braket{C_j}=0$ for $J_4<J_{4c}$.
Nevertheless, the ground state remains on the $1/2$ plateau for $J_4>J_{4c}$ in the thermodynamic limit since $\frac{M}{M_s}=\frac{1}{2}(1+\frac{1}{L})\to\frac{1}{2}$.

Thus far, we have developed the effective field theory to understand the topological quantum phase transition of the spin-1 chain \eqref{H_4} on the $1/2$ magnetization plateau.
We employed the effective field theory approach because the pseudospin approximation seemed powerless.
Still, the weakly-coupled dimer picture at $|\alpha|\ll 1$ gives us some intuition to understand the abrupt charge jump at $J_{4}=J_{4c}$.
When $J_4=\alpha=0$, the ground state is the product state of $\ket{t_1}$.
The $J_4$ interaction weakens antiferromagnetic intradimer coupling, $J-J_4$, and invites the spin-2 state, $\ket{q_2}$, to join the low-energy physics.
Let us compare the energies of two product states $\ket{t_1}_1\ket{t_1}_2\cdots\ket{t_1}_{L/2}$ and $\ket{t_1}_1\ket{t_1}_2\cdots \ket{q_2}_{r_1} \cdots \ket{t_1}_{L/2}$, where we replace $\ket{t_1}_{r_1}$ by $\ket{q_2}_{r_1}$ for one dimer at $r_1$.
Let $E$ and $E'$ be eigenenergies of the former and latter states for $\alpha=0$.
The latter becomes the ground state when their energy difference $\Delta E:=E'-E$,
\begin{align}
    \Delta E &= 2(J-J_4)-h_u
\end{align}
becomes negative, that is, when $J_4>J_{4d}$ with
\begin{align}
    J_{4d} &= \frac{2J-h_u}{2}.
    \label{J4c}
\end{align}
For $\alpha=0$, the first-order transition occurs at $J_4=J_{4d}$ because every $\ket{t_1}$ is replaced by $\ket{q_2}$ as $J_4$ passes $J_{4d}$.
For $\alpha\not=0$, the inter-dimer exchange interaction minimizes the number of $\ket{q_2}$ to minimize the energy cost due to the antiferromagnetic $J\alpha$ interaction.
The number of $\ket{q_2}$ is zero for $J_4<J_{4d}$ and will be one for $J_4>J_{4d}$.
For the parameter set of Fig.~\ref{fig:bondalt_edge}, we obtain
$J_{4d}=0.25$, close enough to $J_{4c}\approx0.295$ where the topological transition actually occurs.
When one of $\ket{t_1}$ is replaced by the spin-2 $\ket{q_2}$, the spin-$2$ object will be delocalized to minimize the energy cost arising from the antiferromagnetic exchange interactions.
If $\ket{q_2}$ carrying the charge one is shunted off to the edges, it is fractionalized to two $1/2$ charges to keep the exact $\mathcal{I}_b$ symmetry.
We thus end up with the edge magnetizations $M_{\rm left}^z=M_{\rm right}^z=1/2$.

\section{Edge magnetization of ferrimagnets}
\label{sec:ferri}

The edge magnetizations hitherto considered are triggered by spatially nonuniform interactions.
Even if the edge magnetization has a topological origin such as the spin-1 Haldane phase of the uniform spin-1 HAFM chain, the staggered magnetic field is necessary to make the edge magnetization visible by lifting the degeneracy of the edge state.
In this sense, we thus far needed spatially nonuniform interactions to trigger the edge magnetization by breaking the inversion symmetry that protects the edge-state degeneracy.

\begin{figure}[t!]
    \centering
    \includegraphics[bb = 0 0 850 550, width=\linewidth]{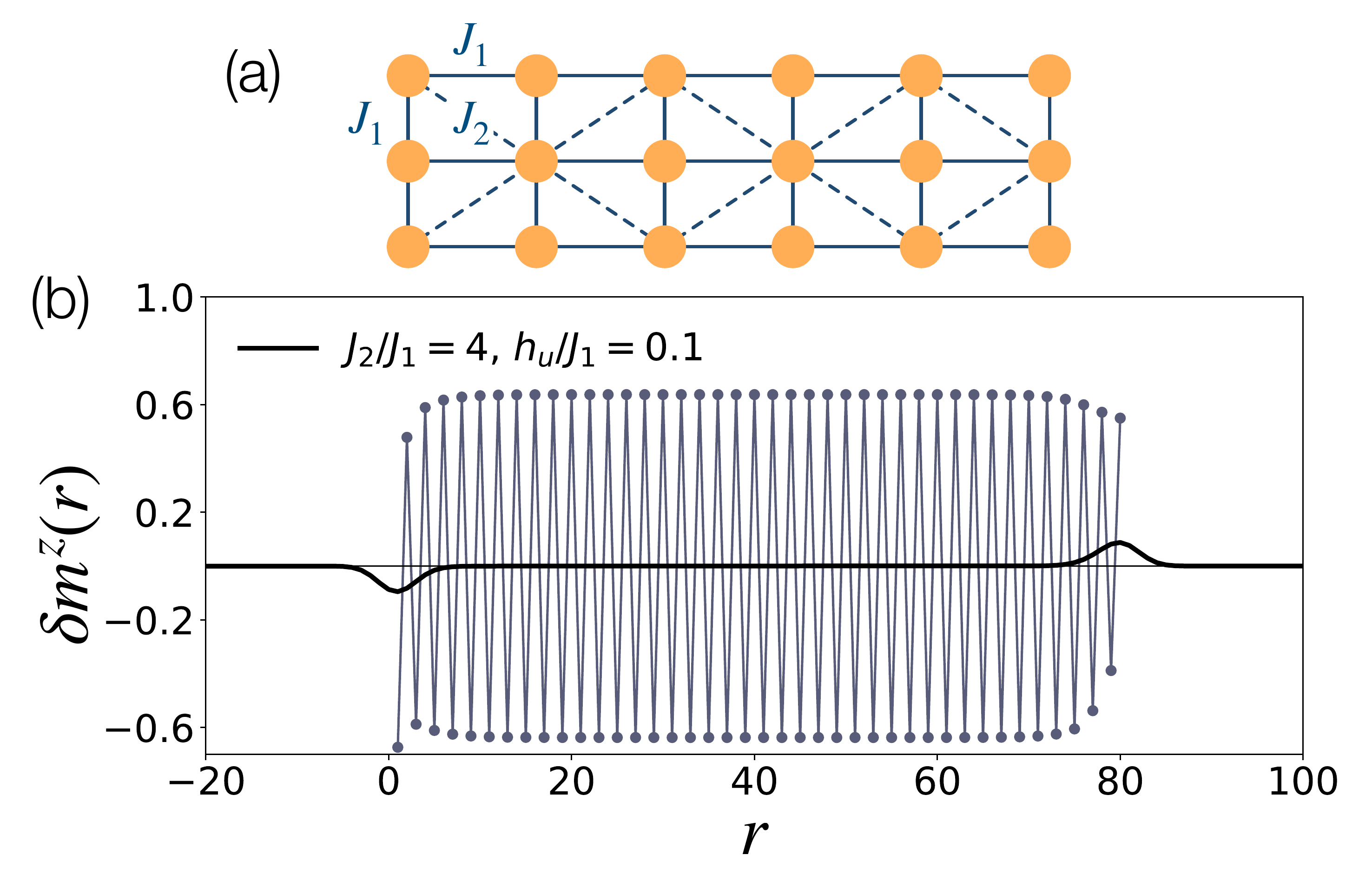}
    \caption{(a) Union-jack strip. (b) Site $r$ dependence of $\braket{C_j}$ and $\delta m^z(r)$ with $N=3$ and $m=1/6$.
    The edge magnetizations are quantized as $M_{\mathrm{right}}^z=-M_{\mathrm{left}}^z=0.5000000$. We used $J_1=1$, $J_4=4$, $h_u=0.1$, and $3L=240$.}
    \label{fig:unionjack}
\end{figure}

Here, we discuss a contrasting case that the uniform magnetic field triggers the edge magnetization by lifting the ground-state degeneracy.
We deal with a spin-$1/2$ HAFM model on a union-jack strip [Fig.~\ref{fig:unionjack}~(a)]~\cite{shimokawa_uj,furuya_uj},
\begin{align}
    \mathcal{H}_{\rm UJ}
    &= J_1\sum_{j=1}^L\sum_{n=1}^3 \bm S_{j,n} \cdot \bm S_{j+1,n} + J_1 \sum_{j=1}^{L}\bm S_{j,2}\cdot (\bm S_{j,1}+\bm S_{j,3})
    \notag \\
    &+J_2\sum_{j=2}^{L-1}\bm S_{j,2}\cdot (\bm S_{j-1,1}+\bm S_{j-1,3} +\bm S_{j+1,1}+\bm S_{j+1,3}),
    \label{H_UJ}
\end{align}
where $\bm S_{j,n}$  is the spin-$1/2$ operator.
The first term denotes the intra-chain interaction, the second term denotes the rung interaction, and the last one denotes the inter-chain diagonal interactions.

For large enough $J_2/J_1>0$, this frustrated three-leg spin ladder exhibits a spontaneous magnetization plateau with $|M|/M_s=1/3$ by spontaneously breaking the SU(2) spin-rotation symmetry~\cite{shimokawa_uj}.
The spontaneous ferromagnetic order is accompanied by an antiferromagnetic order thanks to the lattice structure.
Namely, the ground state has the spontaneous long-range commensurate ferrimagnetic order.
The ferrimagnetic order leads to $\mathcal P\not=0$, as we show below. 
However, similarly to the spin-1 Haldane phase, the edge magnetization is concealed by a $\mathbb Z_2$ symmetry, for example, a $\pi$ rotation symmetry $(S_{j,n}^x,S_{j,n}^y,S_{j,n}^z) \to (S_{j,n}^x,-S_{j,n}^y,-S_{j,n}^z)$ around the $x$ axis.

Different from the spin-1 Haldane phase, an infinitesimal \emph{uniform} magnetic field completely lifts the ground-state degeneracy by choosing one of the spontaneous ferrimagnetic states.
In the presence of the weak uniform magnetic field, the effective field theory of the union-jack strip on the $1/3$ plateau has the following Hamiltonian (Appendix~\ref{app:unionjack}),
\begin{align}
    \mathcal{H}_{\mathrm{UJ}} &= \int_0^L dx\biggl[\frac{v}{2\pi K}(\partial_\mu \phi)^2 +\zeta\cos(2\phi)
    \biggr].
    \label{H_UJ_phi}
\end{align}
The $\phi$ field is related to the charge as $C_j=\sum_{n=1}^3 S_{j,n}^z= \partial_x\phi/\pi$.
The cosine interaction locks $\phi$ to $\bar\phi=\pm\pi/2$ and gives $\mathcal{P}=\mp1/2$.
Tracing the hitherto developed argument, we can confirm that the U(1) and $\mathcal{I}_s$ symmetries protect the quantization of the edge magnetization $\mathcal{P}=1/2$.
Our $3L=240$-site calculation shows the fine quantization $M_{\mathrm{right}}^z=-M_{\mathrm{left}}^z=0.50000000$.
Note that not $\mathcal I_b$ but $\mathcal I_s$ protects the quantization in contrast to the spin-1 chain \eqref{H_4} on the $1/2$ plateau.
This symmetry difference ultimately comes from the difference in number of spin ladders' legs~\footnote{The effect of the number of legs on the inversion symmetries is well exemplified by the $N$-leg spin-$1/2$ ladder at $m=0$~\cite{fuji_spt_eft}. The one-site translation along the leg leads to $\phi(x) \to \phi(x)+\frac{\pi N}{2}$. 
The $\phi$ field is a uniform summation of the $\phi_n$ field on $n$th leg for $n=1,2, \cdots, N$.
Since each $\phi_n$ admits the $\pi/2$ shift by $T_1$, their summation $\phi=\phi_1+\phi_2+\cdots+\phi_N$ admits the $\pi N/2$ shift.
Since $\mathcal I_b=T_1\mathcal I_s$, the $N$ dependence of $T_1$ affects the field-theoretical representation of $\mathcal I_s$ and $\mathcal I_b$. One can find a similar effect in effective field theories on the magnetization plateaus dealt with in this paper.
}.

In the absence of the uniform magnetic field, the magnetization plateau with $m=1/3$ and $m=-1/3$ are degenerate, where the edge magnetizations are concealed by the ground-state degeneracy.
The uniform magnetic field chooses, for example, the $m=1/3$ state and makes the ground state unique and gapped.
Then, nothing conceals the edge magnetization any longer [Fig.~\ref{fig:unionjack}~(b)].

\section{Conclusion and outlook}
\label{sec:conclusions}

\begin{table}[t!]
    \centering
    \begin{tabular}{ccc} \hline \hline
         quantum phase & edge magnetization? & on plateau?  \\ \hline
         odd-spin Haldane (SPT) & no & yes~\cite{oya,takayoshi_laser_curve}
          \\
         induced N\'eel (SPt) & yes & not found yet \\
         large-$D$ (trivial) & no & yes~\cite{sakai_3/2_plateau,kitazawa_3/2_plateau} \\
         quantized $\mathcal P$ & yes & yes \\ \hline \hline
    \end{tabular}
    \caption{Comparison of quantum phases with symmetry-protected quantized edge magnetization $\mathcal P$ (fifth row) with odd-spin Haldane phases (second row), an SPt phase (third row), and a trivial phase (fourth row).
    We numerically and field-theoretically found quantum phases accompanied by edge magnetizations on magnetization plateaus $m=0$ or $m\not=0$, which are referred to as "quantized $\mathcal P$ in the table.
    The odd-spin Haldane phase, a symmetry-protected topological (SPT) phase, exhibit no edge magnetizations because a protecting symmetry forbids it.
    The quantized-$\mathcal P$ phase as well as the odd-spin Haldane phase are realized on a nonzero magnetization plateau~\cite{takayoshi_plateau_eft}.
    The SPt phase can exhibit the edge magnetization but is not found yet on nonzero magnetization plateaus.
    }
    \label{tab}
\end{table}

We discussed the edge magnetization as the magnetic analog of the surface electric charge by using the low-energy effective field theory and the numerical density-matrix renormalization group method.
Low-energy physics of one-dimensional quantum spin systems with or without the magnetization per site is described by the same effective field theory, the sine-Gordon theory.

The sine-Gordon theory is the strongly interacting field theory of the U(1) boson field $\phi$.
We showed that the edge magnetization as the surface electric charge is the zero mode of $\phi$ [Eq.~\eqref{P_phi}].
The quantization of the zero mode is protected by the U(1) spin-rotation and inversion symmetries.
The inversion symmetry can be either the site-centered or bond-centered one, depending on the carrier of the charge [Eq.~\eqref{U_def_plateau}] and the number of spin chains.

We characterized quantum phases of one-dimensional quantum spin systems based on the edge magnetization and the symmetry protection of its quantization.
We found some affinities and differences of this characterization with the odd-spin Haldane phase (a symmetry-protected topological phase) and the SPt phase as summarized in Table.~\ref{tab}.
The edge magnetization turned out to give us an interesting viewpoint of the classification of quantum phases.
Moreover, in principle, the edge magnetization is an observable quantity and will be relevant to experimental studies.

Our field-theoretical results on one-dimensional quantum spin systems will be useful as building blocks to construct magnetic analog of corner magnetizations in two- or three-dimensional quantum spin systems~\cite{watanabe_corner} in the spirit of the coupled-wire construction~\cite{kane_cwc_2002,kane_cwc_2014,meng_cwc_csl,lecheminant_cwc_na_csl}.

\section*{Acknowledgments}

This work is by a Grant-in-Aid for Scientific Research on Innovative Areas ”Quantum Liquid Crystals” Grant No. JP19H05825 (for S.C.F. and M.S.), JSPS KAKENHI No. JP20K03769 (for S.C.F.), and JSPS KAKENHI Grant Nos. JP17K05513 and JP20H01830 (for M.S.).

\appendix

\section{Open boundary condition in spin chains}
\label{app:obc}

The open boundary condition (OBC) on the quantum spin-1/2 chain is formulated in a fermion language~\cite{eggert_obc}.
The spin-1/2 operator $\bm S_j$ is written as~\cite{giamarchi_book}
\begin{align}
    S_j^z &= \psi_j^\dag \psi_j - \frac 12, \\
    S_j^+ &= \psi_j^\dag \exp\biggl(i\sum_{k=1}^{j-1} \psi_k^\dag \psi_k \biggr),
\end{align}
where $\psi_j$ is an annihilation operator of a spinless fermion.
The spinless fermion $\psi(x)=\psi_j/\sqrt{a_0}$ is split into right-moving $\psi_R(x)$ and left-moving $\psi_L(x)$ parts: $\psi(x) = e^{-ik_Fx} \psi_L (x) + e^{ik_F x} \psi_R(x)$
with $x=ja_0$ and the Fermi wavenumber $k_F=\pi/2a_0$~\cite{giamarchi_book}.
We impose the OBC at $x=0$ on the spinless fermion by requiring the following conditions~\cite{eggert_obc},
\begin{align}
    \psi(0) = \psi(L)=0.
    \label{obc_psi}
\end{align}
Note that chiral fermion operaetors, $\psi_{L}$ and $\psi_{R}$ cancel each other so that Eq.~\eqref{obc_psi} holds.
This boundary condition is further translated into that for $\phi(x)$ and $\theta(x)$ via the following bosonization formula~\cite{giamarchi_book},
\begin{align}
    \psi_R(x) &\sim e^{-i(\theta-\phi)}, \qquad \psi_L(x) \sim e^{-i(\theta+\phi)}.
\end{align}
Two boson fields $\phi(x)$ and $\theta(x)$ satisfy the commutation relation,
\begin{align}
    [\phi(x), \theta(y)] = i\pi \Theta_{\rm step}(y-x),
    \label{comm_step}
\end{align}
where $\Theta_{\rm step}(z)$ is the step function,
\begin{align}
    \Theta_{\rm step}(z) &= \left\{
    \begin{array}{ccc}
    1 & & (z>0) \\
    1/2 & & (z=0) \\
    0 & & (z<0)
    \end{array}
    \right..
\end{align}
This bosonization formula leads to, for instance, $(-1)^j S_j^z \approx a_1\sin(2\phi) + \cdots$ because $S_j^z=\psi_j^\dag \psi_j - \frac 12$,
\begin{align}
    \psi^\dag(x)\psi(x)
    &= \psi_R^\dag(x) \psi_R(x) + \psi_L^\dag (x) \psi_L(x)
    \notag \\
    &\quad + (-1)^j \bigl[\psi_R^\dag (x) \psi_L(x)+\psi_L^\dag (x)\psi_R(x)\bigr],
\end{align}
and
\begin{align}
    &\psi_R^\dag (x) \psi_L(x)+\psi_L^\dag (x)\psi_R(x)
    \notag \\
    &\sim  e^{-2i\phi(x) + [\phi(x),\theta(x)]} + e^{2i\phi(x) -[\phi(x),\theta(x)]}
    \notag \\
    &= 2\sin(2\phi).
\end{align}

The boundary condition, $\psi(0)=0$, on the left edge leads to~\cite{eggert_obc}
\begin{align}
    1-e^{2i\phi(0)}=0,
\end{align}
namely, 
\begin{align}
    \phi(0) &= 0 \mod \pi.
    \label{phi_0_0}
\end{align}
On the other edge, we obtain
\begin{align}
    1- e^{2ik_F L} e^{2i\phi(L)}=0.
\end{align}
Since $L/a_0$ must be an even integer to meet the charge-neutrality condition, we find
\begin{align}
    \phi(L) = 0 \mod \pi.
    \label{phi_L_0}
\end{align}
This boundary condition $\phi(0)=\phi(L)=0 \mod \pi$ is consistent with the charge neutrality condition,
\begin{align}
    \sum_{j=1}^L S_j^z =\frac{1}{\pi} [\phi(L)-\phi(0)]=0.
\end{align}

The OBC on the spin chain is interpreted as the Dirichlet boundary condition on $\phi(x)$ at $x=0,\,L$.
Equations~\eqref{phi_0_0} and \eqref{phi_L_0} leads to
\begin{align}
    \partial_t\phi(x)\Bigr|_{x=0,L}=0.
\end{align}
The $\phi$'s canonical conjugate, $\theta$, then satisfies the Neumann boundary condition~\cite{furuya_bbs}:
\begin{align}
    \partial_x\theta(x)\Bigr|_{x=0,L}=0.
\end{align}

\section{Semiclassical bosonization at zero magnetic fields}
\label{app:m=0}

This section describes a derivation of the sine-Gordon theories for the spin-$S$ chain from the O(3) nonlinear sigma model (NL$\sigma$M).
Here, we deal with zero-field cases.

\subsection{Classical Hamiltonian of nonlinear sigma model}

We start with the mapping of the spin operator $\bm S_j$ to slowly varying fields $\bm n(x_j)$ and $\bm L(x_j)$~\cite{sachdev_textbook}:
\begin{align}
    \bm S_j &= S\bm \Omega (x_j),
    \label{S2Omega} \\
    \bm \Omega (x_j) &=  (-1)^j \bm n(x_j)\sqrt{1-\biggl(\frac{a_0\bm L(x_j)}{S}\biggr)^2} + \frac{a_0}S \bm L(x_j),
    \label{Omega_def}
\end{align}
where $S$ is the spin quantum number, $x_j = ja_0$ is the spatial coordinate, and $\bm \Omega(x_j)$ is the three-component unit vector with $|\bm \Omega(x_j)|^2=1$.
Two quantum fields $\bm n(x)$ and $\bm L(x)$ satisfy $|\bm n(x)|^2 = 1$ and $\bm n(x) \cdot \bm L(x) = 0$ for every $x$ so that $|\bm \Omega(x)|^2=1$.
To respect the SU(2) commutation relation $[S_j^a, S_k^b] = i\varepsilon^{abc} \delta_{j,k} S_j^c$, the following commutation relations are required~\cite{sachdev_textbook}.
\begin{align}
    [L^a(x), L^b(x)] &= i\varepsilon^{abc} L^c(x) \delta(x-y), \\
    [L^a(x), n^b(x)] &= i\varepsilon^{abc}  n^c(x) \delta(x-y), \\
    [n^a(x), n^b(x)] &= 0,
\end{align}
where $\varepsilon^{abc}$ is the complete antisymmetric tensor with $\varepsilon^{xyz}=1$ and $\delta_{i,j}$ is the Kronecker's delta.

Let us consider the partition function $Z$ of the spin-$S$ Heisenberg antiferromagnetic spin chain, 
\begin{align}
    \mathcal H = J \sum_j \bm S_j \cdot \bm S_{j+1}.
    \label{H_chain_SIA}
\end{align}
Note that our arguments also applies to $N$-leg spin-$S$ ladders and other related one-dimensional systems~\cite{Affleck_review_1989,senechal1995_ladder,Sierra1996_ladder,dellaringa1997_ladder,sato2007_ladder}.
For simplicity, we take the simplest example \eqref{H_chain_SIA} here.
Performing the Taylor expansion on the exchange interaction $\bm S_j \cdot \bm S_{j+1}$ up to the $O({a_0}^2)$ terms, we obtain
\begin{align}
    \sum_j \bm S_j \cdot \bm S_{j+1}
    &\approx \int \frac{dx}{a_0} \biggl( \frac{S^2{a_0}^2}{2} (\partial_x \bm n)^2  + 2{a_0}^2 \bm L^2 \biggr) + \mathrm{const.}
\end{align}
The effective Hamiltonian is given by 
\begin{align}
    \mathcal H_{\rm cl} &= \int dx \biggl( \frac{gv}{2}\bm L^2 + \frac{v}{2g}(\partial_x \bm n)^2 \biggr),
    \label{H_cl}
\end{align}
with $g=2/S$ and $v=2JSa_0$.
Equation~\eqref{H_cl} represents the classical Hamiltonian in the path integral formalism.
In other words, the Berry phase is yet to be included.
We can express the uniform component $\bm L(x_j)$ of the spin operator $\bm S_j$ in terms of the staggered one, $\bm n(x_j)$.
The Heisenberg equation of motion $\partial_t \bm n = i[\mathcal H_{\rm cl}, \bm n]$ tells us that $\partial_t \bm n = gv \bm L \times \bm n$.
This relation immediately leads to
\begin{align}
    \bm L &= \frac 1{gv} \bm n \times \partial_t \bm n .
    \label{L2n}
\end{align}

\subsection{Berry phase}

The partition function $Z$ of the spin chain \eqref{H_chain_SIA} is written as
\begin{align}
    Z &= \int \mathcal D \bm \Omega  \delta(|\bm \Omega|^2-1) e^{-\mathcal S},
    \label{partition_func}
    \\
    \mathcal S &= \mathcal S_{\rm BP} +\mathcal S_{\rm cl},
    \label{action_total}
\end{align}
in the path-integral formalism, where $\mathcal S$ is the total action and
$\mathcal S_{\rm cl} = \int_0^\beta d\tau \, \mathcal H_{\rm cl}$ is the classical action.
$\beta=1/k_BT$ is the inverse temperature, eventually set to $\beta \to + \infty$.
The other part $\mathcal S_{\rm BP}$ of the action is the Berry phase~\cite{sachdev_textbook,auerbach_textbook}:
\begin{align}
    \mathcal S_{\rm BP}
    &= -iS \sum_j \omega[\bm \Omega(x_j,\tau)],
    \label{BP_def} \\
    \omega[\bm \Omega(x_j,\tau)]
    &= \int_0^\beta d\tau \, \Bigl(1-\cos \gamma(x_j,\tau) \Bigr) \partial_\tau\theta(x_j,\tau),
\end{align}
where $\gamma(x,\tau)$ and $\theta(x,\tau)$ are the polar and azimuthal angles, respectively:
\begin{align}
     \bm \Omega(x,\tau) &= 
     \begin{pmatrix}
     \sin\gamma(x,\tau) \cos\theta(x,\tau) \\ \sin\gamma (x,\tau) \sin\theta(x,\tau) \\ \cos \gamma(x,\tau)
     \end{pmatrix}.
     \label{Omega_polar}
\end{align}
We employed this notation for the angles to make contact with the conventional notation of the Abelian bosonization~\cite{giamarchi_book} in Sec.~\ref{app:bosonization_m=0}.

The Berry phase \eqref{BP_def} gives rise to the well-publicized theta term that determines the ground state's fate in quantum spin chains~\cite{haldane_1983a, haldane_1983b, Affleck_review_1989}.
Note that the classical ground state of the model \eqref{H_chain_SIA} is the N\'eel ordered state.
We can regard $(-1)^j\bm n$ in Eq.~\eqref{Omega_def} as the classical configuration,
\begin{align}
    \bm \Omega(x_j) = (-1)^j \bm n(x_j),
    \label{Omega_cl_Neel}
\end{align}
and $a_0\bm L/S_0$ as its quantum fluctuations.
Let us evaluate the Berry phase \eqref{BP_def} for the classical configuration \eqref{Omega_cl_Neel}.
The Berry phase then becomes
\begin{align}
    \mathcal S_{\rm BP}
    &= -iS \sum_j (-1)^j \omega [\bm n(x_j,\tau)]
    \notag \\
    &= -iS \sum_{j'} \Bigl\{\omega[\bm n(x_{2j'}, \tau)] - \omega[\bm n(x_{2j'-1},\tau)] \Bigr\}
    \notag \\
    &= -iS \int_0^\beta d\tau  \int \frac{dx}2 \frac{\delta \omega[\bm n(x), \tau]}{\delta \bm n(x,\tau)} \cdot \partial_x \bm n(x,\tau).
\end{align}
The functional derivative $\delta \omega[\bm n]/\delta \bm n$ has a simple representation~\cite{auerbach_textbook},
\begin{align}
    \frac{\delta\omega [\bm n]}{\delta \bm n} = \bm n \times \partial_\tau \bm n.
\end{align}
Then, the Berry phase turns into the theta term,
\begin{align}
    \mathcal S_{\rm BP}
    &= i\Theta \int \frac{d\tau dx}{4\pi} \bm n \cdot \partial_\tau \bm n \times \partial_x \bm n,
    \label{BP_theta}
\end{align}
with $\Theta = 2\pi S$.
Inclusion of the quantum fluctuation $a_0\bm L/S$ has no impact on the value of the Berry phase \eqref{BP_theta} since the local modifications of $\bm \Omega$ keep the topological term such as the Berry phase intact.

\subsection{Dual transformation to sine-Gordon theory}

It is well known that the O(3) NL$\sigma$M at zero magnetic fields can be mapped to the sine-Gordon model~\cite{affleck_meron}.
Here, we derive the dual transformation at the operator level.
That is, we relate the boson fields of the sine-Gordon theory to the sigma field $\bm n$ of the O(3) NL$\sigma$M, and ultimately to the original spin.
We call this bosonization formula a ``semiclassical'' bosonization since the O(3) NL$\sigma$M is a semiclassical field theory.
Interestingly, the resultant ``bosonization'' formulas resemble the well-known Abelian bosonization formulas of quantum spins~\cite{giamarchi_book}, as we show later.

To bridge the O(3) NL$\sigma$M and the sine-Gordon model, we add a local interaction, $D(n^z)^2$ with $D>0$, to the classical Hamiltonian \eqref{H_cl}.
This term is akin to the single-ion anisotropy term $D\sum_j (S_j^z)^2$ and the easy-plane exchange anisotropy $-D\sum_j S_j^z S_{j+1}^z$.
The introduction of the easy-plane anisotropy to the spin chain \eqref{H_chain_SIA} does not immediately induce any quantum phase transition regardless of the spin quantum number~\cite{giamarchi_book, chen_spin-1_gspd,tonegawa_spin-2}.

\subsubsection{Action of dual quantum field theory}

The low-energy excitations of the O(3) NL$\sigma$M carry the topological number,
\begin{align}
    Q_{\rm m} = \frac{1}{4\pi} \int d\tau dx \, \bm n \cdot \partial_\tau \bm n \times \partial_x \bm n.
    \label{q_theta}
\end{align}
The topological number \eqref{q_theta} is called the skyrmion number.
The magnetic skyrmion carries $Q_{\rm m} \in \mathbb Z$.
The skyrmion can be split into two merons (Fig.~\ref{fig:meron})~\cite{gross_meron, affleck_meron}, which plays the essential role in what follows.

The meron with $Q_{\rm m}=\pm 1/2$ resembles a vortex with the vorticity $\pm 1$.
The meron avoids the energy cost due to the anisotropy by mostly lying down on the $xy$ plane.
Unlike the vortex with the singular point at its center, the meron avoids the singularity by pointing toward the $z$ axis in a finite spacetime area.
Let us call this area the core.
The core size is a decreasing function of $D$.

\begin{figure}[t!]
    \centering
    \includegraphics[bb = 0 0 1000 600, width=\linewidth]{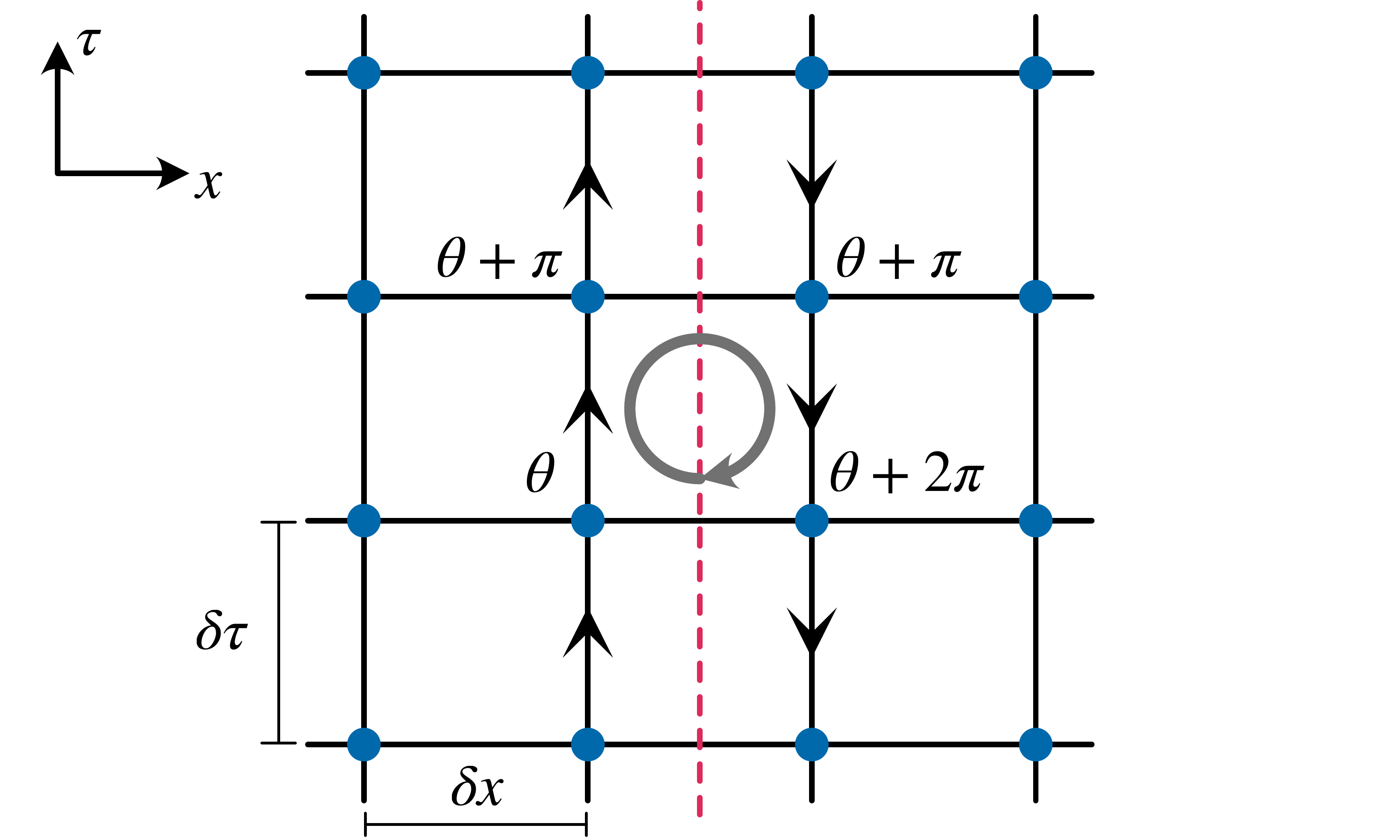}
    \caption{A vortex with $\nu=1$ resting a bond connecting two points $x_j$ and $x_{j+1}$ is shown. The spacetime is discretized to the rectangular lattice.
    The vorticity is defined on a rectangular plaquette indicated by a gray circle with an arrow.
    The $\theta$ field changes by $\pi$ along the vertical line with arrows.
    The vorticity on the plaquette is $\nu = [\theta(x,\tau+\delta \tau) - \theta(x,\tau) - \theta(x+\delta x, \tau+\delta \tau) + \theta(x+\delta x,\tau+\delta \tau)]/2\pi=1$.}
    \label{fig:sq}
\end{figure}

The meron's topological charge \eqref{q_theta} is characterized by the two integers, $(\sigma, \nu)$, where $\sigma =\pm 1$ is the sign of $n^z$ at the center of the meron's core and $\nu \in \mathbb Z$ is the vorticity density.
In what follows, we discretize the ($1+1$)-dimensional spacetime as the rectangular lattice with the lattice spacings $\delta \tau$ and $\delta x$ in the $\tau$ and $x$ directions, respectively (Fig.~\ref{fig:sq}).
If we take $\delta\tau$ and $\delta x$ much larger than the core size of the meron, the meron on the discretized spacetime behaves just like the vortex except for the topological term.
The topological charge \eqref{q_theta} recalls the orientation $\sigma=\pm 1$ of $n^z$ at the core center of the meron.
When the system has $N_{\mathrm{m}}$ merons with $(\sigma_n, \nu_n)$ for $n=1,2,\cdots, N_{\mathrm{m}}$, the net topological charge \eqref{q_theta} is written as~\cite{nagaosa_skyrmion_review}
\begin{align}
    Q_{\rm m} = \sum_{n=1}^{N_{\mathrm{m}}} \frac 12 \sigma_n \nu_n.
    \label{Q_m}
\end{align}
The vorticity density is defined as
\begin{align}
    \nu_n &= \frac{\delta\tau \delta x}{2\pi}(\partial_\tau \partial_x-\partial_x\partial_\tau) \theta(x,\tau).
\end{align}
The net vorticity over the system, $Q_{\rm v}$, is given by
\begin{align}
    Q_{\rm v} &= \sum_n \nu_n =  \frac{1}{2\pi}\int d\tau dx \, (\partial_\tau \partial_x - \partial_x \partial_\tau) \theta(x,\tau).
    \label{vorticity_def}
\end{align}

We are now ready to derive a dual field theory of the O(3) NL$\sigma$M.
The meron has a characteristic length scale $\ell_{\rm c}$ corresponding to the core size.
If the correlation length, $\ell$, of merons is much longer than $\ell_{\rm c}$, the merons can be effectively regarded as vortices at the length scale  $\gtrsim \ell$.
We can assume $\ell\gg \ell_{\rm c}$ without loss of generality thanks to $D$. Larger $D/v$ shrinks the core size $\ell_{\rm c}$ and expands the correlation length $\ell$ simultaneously.
Accordingly, we can construct the low-energy effective field theory of meron similarly to that for the vortex~\cite{KT_1973, Kosterlitz_1974}.
Following the standard argument of the dual transformation of the two-dimensional XY model to the sine-Gordon theory~\cite{KT_1973, Kosterlitz_1974}, we rewrite the action $\mathcal S$ and the partition function $Z$ as
\begin{align}
    \mathcal S &=   \frac 1{2g} \int d\tau dx\, (\partial_\mu \theta)^2 -(\ln \zeta) \sum_{n=0}^{N_{\rm m}} \nu_n^2+ i\frac{\Theta}{2} \sum_{n=0}^{N_{\rm m}} \sigma_n \nu_n, 
    \label{S_meron} \\
    Z &= \sum_{N_{\rm m}=0}^\infty  \int \mathcal D\mathcal \theta \, \exp(-\mathcal S),
\end{align}
where $(\partial_\mu\theta)^2 = (\partial_x\theta)^2 + (\partial_\tau\theta)^2$ and  $\zeta$ is a fugacity of the meron~\cite{affleck_meron}.
$N_{\rm m}=0,1,2, \cdots$ is the total number of merons.
We set $\nu_0=\sigma_0=0$ for $N_{\mathrm{m}}=0$.
Note that the velocity $v$ is set to unity for simplicity.
We can resurrect $v$ whenever we want.
The second term of Eq.~\eqref{S_meron} represents the energy cost to create the meron with $(\sigma_n, \nu_n)$, which is introduced here based on physical considerations~\cite{Kosterlitz_1974}.
The energy cost is independent of $\sigma_n = \pm 1$ thanks to the global  $\mathbb Z_2$ symmetry under $n^z \to - n^z$.
The easy-plane anisotropy $D(n^z)^2$ is encoded in the fugacity $\zeta$.

Let us introduce an auxiliary field $J_\mu$ for $\mu=\tau,x$ through the Hubbard-Stratonivich transformation~\cite{takayoshi_plateau_eft}.
\begin{align}
    Z &= \sum_{N_{\rm m}} \int \mathcal D\theta \mathcal D J_\tau \mathcal DJ_x   \exp\biggl( -\frac{1}{2\pi K} \int d\tau dx\, {J_\mu}^2 
    \notag \\
    &\quad +\frac{i}{\pi} \int d\tau dx \,J_\mu \partial_\mu \theta + (\ln \zeta) \sum_n \nu_n^2 - i\frac{\Theta}{2} \sum_n \sigma_n \nu_n
    \biggr),
    \label{Z_J_theta}
\end{align}
with $K$ being
\begin{align}
    K &= \frac{\pi}{g}.
    \label{K_def}
\end{align}
Here, we split $\theta$ into a regular part $\theta_{\rm r}$ and a vortex part $\theta_{\rm v}$: $\theta = \theta_{\rm r} + \theta_{\rm v}$.
These two parts are distinguished by the vorticity,
\begin{align}
    (\partial_\tau\partial_x-\partial_x\partial_\tau)\theta_{\rm r}(x,\tau) &=0, \notag \\
    (\partial_\tau\partial_x-\partial_x\partial_\tau)\theta_{\rm v}(x,\tau) &\not=0.
\end{align}
Integrating $\theta_{\rm r}$ field in Eq.~\eqref{Z_J_theta}, we obtain the delta function
\begin{align}
    \int \prod_{\tau,x} \mathcal D\theta_{\rm r}(x,\tau) \exp\biggl(-\frac{i}{\pi} (\partial_\mu J_\mu)\theta_{\rm r} \biggr) \propto \prod_{\tau,x} \delta(\partial_\mu J_\mu).
\end{align}
On the other hand, completing the square with respect to $J_\mu$, we find that the following relation holds along a path with the largest contribution to the path integral:
\begin{align}
    J_\mu = iK\partial_\mu \theta_{\rm v},
\end{align}
for $\mu=\tau, x$.
Hereafter, we denote $\theta_{\rm v}$ as $\theta$ for simplicity except when we stress the difference of $\theta_{\rm v}$ and $\theta$.

The condition, $\partial_\mu J_\mu=0$, imposed by the delta function $\delta(\partial_\mu J_\mu)$ is automatically met if we write the $J_\mu$ field as
\begin{align}
    J_\mu = \varepsilon^{\mu\nu}\partial_\nu \phi,
\end{align}
where $\varepsilon^{\mu\nu}$ is the two-dimensional complete anatisymmetric tensor with $\varepsilon^{\tau x}=1$.
The field $\phi$ introduced so is dual to $K\theta$ in a sense that they satisfy the Cauchy-Riemann relation,
\begin{align}
    \frac Kv \partial_t \theta &=  \partial_x \phi, \\
    -K\partial_x \theta &= \frac 1v \partial_t \phi.
    \label{Cauchy-Riemann}
\end{align}
The $\phi$ field is coupled to the vorticity density through the following term of the action \eqref{Z_J_theta}:
\begin{align}
    \frac i\pi \int d\tau dx \,J_\mu \partial_\mu\theta
    &=-\frac{i}{\pi} \int d\tau dx \, \varepsilon^{\mu\nu} \phi \partial_\nu \partial_\mu \theta.
\end{align}
We derived the right hand side by integrating $\phi$ by parts.
If we discretize the spacetime to a rectangular lattice (Fig.~\ref{fig:sq}), we can further rewrite it as
\begin{align}
    \exp \biggl(\frac{i}{\pi} \int d\tau dx \, J_\mu \partial_\mu \theta \biggr)
    &= \prod_{\tau,x} \exp(2i\phi \nu_n ).
\end{align}
We obtain the following expression of the partition function:
\begin{align}
    Z &= \sum_{N_{\rm m}}  \int \mathcal D \phi  \prod_{\tau,x} \exp\biggl( -\frac{1}{2\pi K} (\partial_\mu \phi)^2 + 2i \phi \nu_n
    \notag \\
    &\quad + (\ln \zeta) \nu_n^2 - i\frac{\Theta}{2}\sigma_n \nu_n \biggr).
\end{align}
Let us keep the $N_{\rm m}=0, 1$ contributions only.

Since the larger $\nu_n$ requires larger energy cost to create the meron, we limit ourselves to $\nu_n = 0$ or $\nu_n = \pm 1$ for $N_{\rm m}=1$.
Then the partition function is approximated as
\begin{align}
    Z &\approx \int \mathcal D \phi \prod_{\tau,x} \exp\biggl( -\frac{1}{2\pi K}  (\partial_\mu \phi)^2\biggr)
    \notag \\
    &\quad \times \biggl[
    1+ \zeta (e^{2i \phi} + e^{-2i\phi}) (e^{-i\frac{\Theta}{2}} + e^{i\frac \Theta 2}) \biggr]
    \notag \\
    &\approx \int \mathcal D \phi \prod_{\tau,x} \exp\biggl( -\frac{1}{2\pi K}  (\partial_\mu \phi)^2 +4\zeta \cos(\Theta/2) \cos(2\phi) \biggr)
    \notag \\
    &= \int \mathcal D \phi \exp(-\mathcal S_{\rm dual}),
\end{align}
with the dual action,
\begin{align}
    \mathcal S_{\rm dual}
    &= \frac{v}{2\pi K} \int d\tau dx \,(\partial_\mu \phi)^2 
    \notag \\
    &\qquad - 4\zeta \cos\biggl(\frac{\Theta}{2}\biggr) \int d\tau dx \, \cos(2\phi).
    \label{S_dual_m=0_2phi}
\end{align}
Since $\Theta=2\pi S$, the dual action \eqref{S_dual_m=0_2phi} is consistent with the existence of the symmetric gapped quantum phase for $S\in \mathbb Z$.
Besides, the sine-Gordon theory \eqref{S_dual_m=0_2phi} shows that the Haldane phases for odd $S$ and even $S$ belong to different phases.
The odd-spin Haldane phase is the symmetry-protected topological phase whereas the even-spin one is topologically trivial~\cite{pollmann_haldane_2012}.

When $S\in\mathbb Z+1/2$, the coupling constant vanishes, $\cos(\Theta/2) = \cos(\pi S) = 0$.
The merons with odd vorticities are then forbidden.
Instead, the merons with even vorticities should be taken into account.
The largest contribution comes from a pair of merons with $\nu_1=\nu_2=\pm 1$
Including these merons, we are led to
\begin{align}
    \mathcal S_{\rm dual} &= \frac{v}{2\pi K} \int d\tau dx \, (\partial_\mu \phi)^2 - 4\zeta^2 \cos \Theta \int d\tau dx \, \cos(4\phi),
    \label{S_dual_m=0_4phi}
\end{align}
for $S\in \mathbb Z+1/2$.
When the ground state described by the sine-Gordon model \eqref{S_dual_m=0_4phi} is gapped, the ground state is doubly degenerate due to the spontaneous breaking of the $\phi \to \phi + \pi/2$ symmetry.
As we see below, this symmetry is the one-site translation symmetry of the spin chain.

\subsubsection{Operator relations between spin and dual boson}\label{app:bosonization_m=0}

To represent spin chain's symmetries in terms of the dual boson field, $\phi$, we need a translation dictionary from the spin to the boson.
Let us recall that the sine-Gordon theories \eqref{S_dual_m=0_2phi} and \eqref{S_dual_m=0_4phi} are derived from the NL$\sigma$M.
Since we already have the translation rules \eqref{S2Omega} and \eqref{Omega_def} from the spin to the sigma field $\bm n$, we only need to establish the translation from the sigma field to the $\phi$ boson.
If we completely ignore the quantum fluctuations, we find $\bm \Omega(x_j) = (-1)^j \bm n(x_j)$.
With the polar coordinate,
\begin{align}
    \bm n(x,\tau)
    &= 
    \begin{pmatrix}
    \sin\gamma(x,\tau) \cos \theta(x,\tau) \\
    \sin\gamma(x,\tau) \sin\theta(x,\tau) \\
    \cos\gamma(x,\tau)
    \end{pmatrix}.
    \label{n_polar}
\end{align}
Note that we are focused on the physics at the length scale much longer than the core size.
Almost everywhere is thus outside the meron's core.
The polar angle is fixed to $\gamma=\pi/2$ outside the core.
At the classical level,
we have $n^x = \cos \theta$ and $n^y = \sin \theta$ but the others, $n^z$, $L^x$, $L^y$, and $L^z$, are zero.
The latter quantities become nonzero when the quantum fluctuation is taken into account.

Previously, we found an equation \eqref{L2n} to relate $\bm L$ to $\bm n$.
For the $z$ component, 
\begin{align}
   a_0L^z &=  \frac{a_0}{gv} (\bm n \times \partial_t \bm n)^z
    \notag \\
    &= \frac{a_0K}{\pi v} \partial_t \theta.
\end{align}
Using the Cauchy-Riemann relation \eqref{Cauchy-Riemann}, we obtain
\begin{align}
    a_0L^z &=  \frac{a_0}{\pi} \partial_x \phi.
\end{align}
The equal-time commutation relation, $[L^z(x), n^a(y)]= i\varepsilon^{zab}n^b(x) \delta(x-y)$ is then rephrased as $[\partial_x \phi(x), e^{i\theta(y)}] = \pi  e^{i\theta(x)}\delta(x-y)$, which implies
\begin{align}
    [\partial_x\phi(x), \theta(y)] &= \pi i \delta(x-y),
\end{align}
or equivalently,
\begin{align}
    [\phi(x), \partial_y \theta(y)] &= \pi i \delta(x-y).
    \label{commutation_relation_phi}
\end{align}
These commutation relations are exactly identical to the canonical one for the U(1) compact boson field of the TL liquid~\cite{giamarchi_book} and also equivalent to Eq.~\eqref{comm_step}.

Next, we look into $n^z$.
This quantity is coupled to the staggered magnetic field through the Zeeman energy, $-h_s\sum_j (-1)^j S_j^z = -(Sh_s/a_0) \int dx \, n^z(x)$.
The staggered field makes merons with $\sigma=\pm$ nonequivalent.
In other words, $h_s$ makes the fugacity dependent on $\sigma_n$: $\zeta_{\sigma_n}$.
We can include the staggered field into the action \eqref{S_meron} as follows.
\begin{align}
    \mathcal S
    &=\frac 1{2g} \int d\tau dx\, (\partial_\mu \theta)^2 - \sum_{n=0}^{N_{\rm m}}(\ln \zeta_{\sigma_n}) \nu_n^2
    \notag \\
    &\qquad + i\frac{\Theta}{2} \sum_{n=0}^{N_{\rm m}} \sigma_n \nu_n.
\end{align}
Repeating the dual transformation, we obtain
\begin{align}
    Z &\approx \int \mathcal D \phi \prod_{\tau,x} \exp\biggl( -\frac{1}{2\pi K}  (\partial_\mu \phi)^2\biggr)
    \notag \\
    &\quad \times \biggl[
    1+ \zeta_+ e^{2\phi i} e^{-i\frac{\Theta}{2}}
    + \zeta_- e^{2\phi i} e^{i\frac{\Theta}{2}}
    \notag \\
    &\quad + \zeta_+ e^{-2\phi i} e^{i\frac{\Theta}{2}}
    +\zeta_- e^{2\phi i} e^{-i\frac{\Theta}{2}}\biggr]
    \notag \\
    &=\int \mathcal D \phi \prod_{\tau,x} \exp\biggl( -\frac{1}{2\pi K}  (\partial_\mu \phi)^2\biggr)
    \notag \\
    &\quad \times \biggl[
    1+ 2\zeta_+ \cos\biggl(2\phi -\frac{\Theta}{2}\biggr)
    +  2\zeta_- \cos\biggl(2\phi +\frac{\Theta}{2}\biggr)
    \biggr]
    \notag \\
    &\approx \int \mathcal D \phi \prod_{\tau,x} \exp\biggl( -\frac{1}{2\pi K}  (\partial_\mu \phi)^2 \biggr)
    \notag \\
    &\quad \times \biggl[ 1+2(\zeta_++\zeta_-) \cos\biggl(\frac{\Theta}{2}\biggr) \cos(2\phi) 
    \notag \\
    &\quad + 2(\zeta_+-\zeta_-) \sin\biggl(\frac{\Theta}{2}\biggr) \sin(2\phi)
    \biggr].
\end{align}
Here, we expand the right hand side about $h_s$.
The expansion of the fugacities $\zeta_\pm = \zeta \pm c_\zeta h_s$ with a constant $c_\zeta$ leads to
\begin{align}
    Z
    &\approx \int \mathcal D \phi \prod_{\tau,x} \exp\biggl(-\frac{1}{2\pi K}(\partial_\mu\phi)^2 \biggr)
    \notag \\
    &\quad \times \biggl[1+ 4\zeta \cos\biggl( \frac{\Theta}{2} \biggr) \cos(2\phi) 
    \notag \\
    &\qquad +4c_\zeta h_s \sin\biggl(\frac{\Theta}{2}\biggr) \sin(2\phi)
    \biggr]
    \notag
    \\
    &\approx \int \mathcal D \phi \, \exp(-\mathcal S_{\rm dual}).
\end{align}
The dual action is thus given by
\begin{align}
    \mathcal S_{\rm dual}
    &= \frac{1}{2\pi K}(\partial_\mu \phi)^2 -4\zeta \cos(\Theta/2) \cos(2\phi)
    \notag \\
    &\quad +a_1h_s\sin(\Theta/2) \sin(2\phi),
\end{align}
with a constant $a_1$.
The last term implies
\begin{align}
    n^z &=a_1\sin(\Theta/2) \sin(2\phi).
    \label{nz2phi}
\end{align}

Finally, we rewrite $L^x$ and $L^y$.
\begin{align}
    L^x
    &= \frac 1{gv} (n^y\partial_t n^z-n^z\partial_tn^x)
    \notag \\
    &= \frac{a_1}{gv} \sin(\Theta/2) \bigl\{\sin\theta (\partial_t\phi) \cos(2\phi) \notag \\
    &\qquad -\sin(2\phi) (\partial_t\theta) \cos\theta
    \bigr\},
    \\
    L^y 
    &= \frac 1{gv} (n^z \partial_t n^x 
    - n^x \partial_t n^z)
    \notag\\
    &= -\frac{a_1}{gv} \sin(\Theta/2) \bigl\{- \sin(2\phi) (\partial_t\theta) \sin\theta
    \notag \\
    &\qquad -2\cos \theta (\partial_t\phi) \cos(2\phi)
    \bigr\}.
\end{align}
The operator-product expansion of the $\phi$ field~\cite{francesco_yellow},
\begin{align}
    \partial_t \phi(x+a_0) \cos(2\phi(x))
    &= \frac{1}{ia_0} \sin(2\phi(x)) + \cdots,
\end{align}
and a similar expansion for $\theta$ lead to
\begin{align}
    L^x &= -i b_1 \sin(\Theta/2) \sin \theta \sin (2\phi),
    \label{Lx2phi} \\
    L^y &= i b_1 \sin(\Theta/2) \cos \theta \sin(2\phi),
    \label{Ly2phi}
\end{align}
with a constant $b_1 \in \mathbb R$.

We could finally translate the spin operator into the $\phi$ and $\theta$ terms.
\begin{align}
    S_j^z &=  \frac{a_0}{\pi}\partial_x \phi + (-1)^j a_1 \sin(\pi S) \sin(2\phi),
    \label{Sz2phi_m=0} \\
    S_j^+ &= e^{i\theta} \bigl[(-1)^j +b_1 \sin(\pi S)\sin(2\phi) \bigr].
    \label{S+2phi_m=0}
\end{align}
The dimer order parameter $(-1)^j \bm S_j \cdot \bm S_{j+1}$ is bosonized as
\begin{align}
    (-1)^j \bm S_j \cdot \bm S_{j+1} = d\cos(2\phi) + \cdots
    \label{dimer_bosonized}
\end{align}
with a nonuniversal constant $d\in \mathbb R$~\cite{Orignac2004_ope_dimer,takayoshi_coeff,hikihara_coeff_dimer,berg_boson_string}.
This bosonization formula follows from operator product expansions such as
\begin{align}
    (-1)^j S_j^z S_{j+1}^z &\approx \frac{2a_0a_1}{\pi} \partial_x\phi \sin(2\phi(x+a_0)) + \cdots
    \notag \\
    &\sim \cos(2\phi(x)) + \cdots.
\end{align}

Surprisingly,  Eqs.~\eqref{Sz2phi_m=0}, \eqref{S+2phi_m=0}, and \eqref{dimer_bosonized} are identical to the standard bosonizaton formulas~\cite{giamarchi_book} for $S=1/2$.
We call Eqs.~\eqref{Sz2phi_m=0} and \eqref{S+2phi_m=0}  semiclassical bosonization formulas.
The relations \eqref{Sz2phi_m=0} and \eqref{S+2phi_m=0} imply that $\phi$ and $\theta$ for $S\in \mathbb Z+1/2$  are compactified as
\begin{align}
    \phi \sim \phi + \pi, \qquad \theta \sim \theta+2\pi.
\end{align}

Note that the staggered component of Eq.~\eqref{Sz2phi_m=0} vanishes when $S\in\mathbb Z$.
The staggered component is not absent but represented as $\sin\phi$ for $S\in\mathbb Z$.
This $S$ dependence of the bosonization formulas is related to the LSM theorem~\cite{lsm,furuya_wzw}.
For $S\in \mathbb Z+1/2$, the anisotropy $D(n^z)^2\propto -\cos(4\phi)$ does not induce the unique gapped ground state.
If $\cos(4\phi)$ is relevant and makes the ground state gapped, the ground state breaks the one-site translation symmetry, $\phi \to \phi + \frac{\pi}{2}$, spontaneously, as we see soon later.
For $S\in \mathbb Z$, by contrast, $D(n^z)^2$ with large enough $D>0$ makes the unique gapped ground state, which is the large-$D$ state, $\bigotimes_{j}\ket{S_j^z=0}$.
The effective Hamiltonian \eqref{S_dual_m=0_2phi} implies that $D(n^z)^2 \propto - D\cos(2\phi)$.
Here, the minus sign comes from the fact that the large-$D$ phase is topologically trivial.
We can deduce
\begin{align}
    n^z \propto \sin \phi, \qquad (S\in \mathbb Z),
    \label{nz_int}
\end{align}
for $S\in \mathbb Z$.
Equation~\eqref{nz_int} is consistent with the $S$ dependence of the one-site translation symmetry.
According to Eq.~\eqref{Sz2phi_m=0},
the one-site translation $T_1:\,\bm S_j \to \bm S_{j+1}$ can be rephrased as
\begin{align}
    T_1: \quad \phi(x) \to \phi(x) + \frac \pi 2 \mod \pi, 
    \label{T1_half_int}
\end{align}
for $S\in \mathbb Z+ 1/2$.
On the other hand, $T_1$ should act on $\phi$ for $S\in \mathbb Z$ as
\begin{align}
    T_1: \quad \phi(x) \to \phi(x) + \pi \mod \pi,
    \label{T1_int}
\end{align}
for $S\in \mathbb Z$ because the effective field theory \eqref{H_SG_4} is $T_1$-invariant and also because the integer-spin HAFM chain can be seen as a two-leg HAFM ladder of the half-odd-integer spin~\cite{fuji_spt_eft}.
For $S\in \mathbb Z$, the two boson fields $\phi$ and $\theta$ will be compactified as
\begin{align}
    \phi \sim \phi + \pi, \qquad \theta \sim \theta+2\pi.
\end{align}
This compactification of $\phi$ is consistent with the deduced relation \eqref{nz_int}.
However, unfortunately, no microscopic derivations of Eq.~\eqref{nz_int} are yet available.

In the main text, we discuss the symmetry protection of the quantization of the edge magnetization.
In the spin-$S$ chains,
the protecting symmetries are the U(1) spin-rotation symmetry and the site-centered inversion symmetry.
The rotation around the $z$ axis by an angle $\varphi$ is obviously translated into
\begin{align}
    \phi(x) &\to \phi(x), \qquad
    \theta(x) \to \theta(x) + \varphi.
    \label{U1_phi_m=0}
\end{align}
The site-centered inversion symmetry is 
\begin{align}
    \phi(x) &\to -\phi(L-x) + \pi S, \qquad \theta(x) \to \theta(L-x).
    \label{Is_phi_m=0}
\end{align}
The symmetries \eqref{U1_phi_m=0} and \eqref{Is_phi_m=0} forbid $\cos \theta$, $\sin\theta$, and $\cos(2\phi)$ for the $S\in \mathbb Z+1/2$ case \eqref{H_SG} and forbids $\cos \theta$, $\sin\theta$, and $\sin(2\phi)$ for the $S\in \mathbb Z$ case \eqref{H_SG_cos}.

\section{Semiclassical bosonization on magnetization plateaus}
\label{app:m>0}

\subsection{Classical Hamiltonian of nonlinear sigma model}

Our starting point for $m>0$ is also the relation \eqref{Omega_def}.
For $m=0$,  we first considered the classical N\'eel configuration and took the quantum fluctuation into account later.
On the magnetization plateau with $m>0$, instead, the classical configuration is the transverse N\'eel state with the longitudinal uniform magnetization:
\begin{align}
    \bm \Omega(x_j,\tau)
    &= 
    \begin{pmatrix}
    (-1)^j n^x(x_j,\tau) \sqrt{1-(m/S)^2} \\ 
    (-1)^j n^y(x_j,\tau) \sqrt{1-(m/S)^2} \\
    m/S
    \end{pmatrix}
    \notag \\
    &=
    \begin{pmatrix}
    (-1)^j \sin\gamma_0 \cos\theta(x_j,\tau) \\
    (-1)^j \sin\gamma_0 \sin\theta(x_j,\tau) \\
    \cos\gamma_0
    \end{pmatrix},
    \label{Omega_cl_plateau}
\end{align}
where $\cos\gamma_0 = m/S$ is fixed to a constant.

In Sec.~\ref{app:m=0}, the staggered part $(-1)^j\bm n(x_j)$ of $\bm \Omega(x_j)$ is the classical configuration and the uniform part $(a_0/S)\bm L(x_j)$ is the quantum fluctuation.
In this section, we regard
Eq.~\eqref{Omega_cl_plateau}
as the classical configuration and the other part of $\bm \Omega(x_j,\tau)$,
\begin{align}
    \begin{pmatrix}
     (a_0/S)L^x \\ (a_0/S) L^y \\ (-1)^jn^z(x_j,\tau) \sqrt{1-(m/S)^2}
    \end{pmatrix},
\end{align}
as the quantum fluctuation.

Similarly to Sec.~\ref{app:m=0}, we also consider the spin-$S$ HAFM chain,
\begin{align}
    \mathcal H = J \sum_j \bm S_j \cdot \bm S_{j+1} -h_u \sum_j S_j^z.
\end{align}
The classical Hamiltonian, $\mathcal H_{\rm cl}$, in the path integral formalism is almost identical to Eq.~\eqref{H_cl}:
\begin{align}
    \mathcal H_{\rm cl}
    &= \int \frac{dx}{a_0} \biggl( \frac{JS^2{a_0}^2}{2}(\partial_x\bm n)^2 + 2J{a_0}^2\bm L^2 
    \notag \\
    &\quad -h_u (a_0 L^z +{a_0}^2 \partial_x L^z) \biggr) + \mathrm{const.}
    \notag \\
    &= \int dx \biggl( \frac{v}{2g}(\partial_x\bm n)^2 + \frac{gv}{2}\bm L^2-h_uL^z \biggr)+\mathrm{const.}
\end{align}
The Heisenberg equation of motion for $\bm n$ relates $\bm L$ and $\partial_t\bm n$:
\begin{align}
    \bm L &= \frac{1}{gv} \bm n \times (\partial_t \bm n + \bm h_u \times \bm n),
\end{align}
with $\bm h_u = (0 \> 0 \> h_u)^T$.
The uniform magnetic field $h_u$ is fixed so that $a_0h_u/gv = m$.

\subsection{Berry phase}

At first glance, the classical Hamiltonian $\mathcal H_{\rm cl}$ does not make much difference from that for $m=0$.
By contrast, the Berry phase is completely different.
The value of the  Berry phase is insensitive to the local modifications of $\bm \Omega(x_j,\tau)$ thanks to its topological nature.
Accordingly, the Berry phase is governed by the classical configuration in the semiclassical approach.

In what follows, we rewrite the Berry phase of the classical configuration \eqref{Omega_cl_plateau}.
Note that the following argument just repeats those of Refs.~\cite{lamas_plateau_2011,takayoshi_plateau_eft}.
Still, it is worth repeating here because we are to derive the Berry phase later in similar but more extended situations.

First, we note that the Berry phase for the classical configuration \eqref{Omega_cl_plateau} of $\bm \Omega(x_j)$ is identical to that for another configuration $\tilde{\bm \Omega}(x_j,\tau)$, that is,
\begin{align}
    \tilde{\bm \Omega} (x_j,\tau)
    &= \begin{pmatrix}
    \sin\gamma_0 \cos\theta(x_j,\tau) \\
    \sin\gamma_0 \sin\theta(x_j,\tau) \\
    \cos\gamma_0
    \end{pmatrix},
\end{align}
except for a certain constant.
More precisely~\cite{tanaka_plateau_eft},
\begin{align}
    -iS \sum_j \omega [\bm \Omega(x_j,\tau)] &= -iS\sum_j \omega [\tilde{\bm \Omega}(x_j,\tau)] +i\pi \sum_j (S-m).
\end{align}
Since the second term on the right-hand side is merely a constant, we can identify these two Berry phases.

Next, following Ref.~\cite{takayoshi_plateau_eft}, we rewrite the Berry phase for $\tilde{\bm \Omega}$ as follows.
\begin{widetext}
\begin{align}
    \mathcal S_{\rm BP}
    &= -iS \sum_{j=1}^L \omega [\tilde{\bm \Omega} (x_j,\tau)]
    \notag \\
    &= -iS \sum_{j'=1}^{L/2} \omega [\tilde{\bm \Omega} (x_{2j'},\tau)] -iS \sum_{j'=1}^{L/2} \biggl(
    2(1-\cos\gamma_0 ) \int_0^\beta d\tau \, \partial_\tau \theta(x_{2j'-1},\tau) - \omega [\tilde{\bm \Omega}(x_{2j'-1},\tau)] \biggr)
    \notag \\
    &= -iS \sum_{j=1}^L (-1)^j \omega[\tilde{\bm \Omega}(x_j,\tau)]- 2i(S-m) \sum_{j'=1}^{L/2} \int_0^\beta d\tau\, \partial_\tau \theta(x_{2j'-1},\tau)
    \notag \\
    &= i\pi (S-m) Q_{\rm v} - i(S-m) \int_0^L \frac{dx}{a_0} \int_0^\beta d\tau\, \partial_\tau \theta(x,\tau)
    \label{BP_tOmega}
\end{align}
\end{widetext}
In the last line, we used the one-site translation symmetry.
$Q_{\rm v}=\sum_n\nu_n$ represents the net vorticity on the two-dimensional spacetime, \eqref{vorticity_def}.
We arrived at the following form of the Berry phase.
\begin{align}
    \mathcal S_{\rm BP} &= i\pi (S-m) Q_{\rm v} + \mathcal S_{\rm LSM}
    \label{BP_vortex}, \\
    \mathcal S_{\rm LSM} 
    &= -i(S-m)\int_0^L \frac{dx}{a_0} \int_0^\beta d\tau\, \partial_\tau \theta(x,\tau).
    \label{S_LSM}
\end{align}
As Ref.~\cite{takayoshi_plateau_eft} mentioned, the term \eqref{S_LSM} is related to the LSM theorem~\cite{lsm}.
Let us denote the ground state on the magnetization plateau as $\ket{\psi_0}$.
The ground state has a configuration, $\{\theta(x,\tau)\}_{\{x,\tau\}\in \mathbb R^2}$, of the $\theta$ field.
Here, we consider another state, $\ket{\psi_1}$, with a slowly shifted configuration $\{\theta'(x,\tau)\}_{\{x,\tau\}\in \mathbb R^2}$, related to $\theta(x,\tau)$ through
\begin{align}
    \theta'(x,\tau) &= \theta(x,\tau) - \frac{2\pi a_0}{\beta L}\tau.
    \label{theta'_LSM}
\end{align}
The $\theta'$ field respects the periodic boundary condition only in the $L\to +\infty$ limit since $\theta'(x,\tau+\beta)=\theta'(x,\tau)-2\pi a_0/L$.
Nevertheless, it is useful to consider the shifted field $\theta'$ because it works as a variational configuration of the genuine low-energy state.

The analytical continuation $\tau \to it$ clarifies the physical meaning of the shift \eqref{theta'_LSM}.
The shift $i(2\pi a_0/\beta L)t$ of $\theta$ is an insertion of a gauge field $\bm A=(A_0, A_1)$ with
\begin{align}
    A_0 &=0, \\
    A_1 &= \frac{2\pi a_0}{\beta L} t,
\end{align}
in the real time $t$~\cite{yao_boundary}.
The gauge field is gradually increased from the time $t=0$ to $t=\beta$, that is, $A_1(t=0) = 0$ and $A_1(t=\beta) = 2\pi a_0/L$.
The latter corresponds to the unit flux insertion~\cite{furuya_pol_amp}.
The gauge field is adiabatically inserted in the $\beta \to + \infty$ limit.

The state $\ket{\psi_1}$ has a vanishing excitation energy in the thermodynamic limit at zero temperature because the difference of the classical energies,
\begin{align}
    \mathcal H_{\rm cl}[\theta']-\mathcal H_{\rm cl}[\theta]
    &= -\frac{2vK}{\beta L} \int_0^L dx \, \partial_x \theta(x,\tau) 
    \notag \\
    &\qquad +\frac{L}{a_0} \frac{vK}{2\pi} \biggl( \frac{2\pi a_0}{\beta L}\biggr)^2,
\end{align}
is vanishing in the limit of $L \to + \infty$.

However, the shift \eqref{theta'_LSM} affects the LSM term.
\begin{align}
    \mathcal S_{\rm LSM}
    &=-i(S-m) \int_0^L\frac{dx}{a_0} \int_0^\beta d\tau \, \partial_\tau \theta'(x,\tau)
    \notag \\
    &= -i(S-m) \int_0^L \frac{dx}{a_0} \int_0^\beta d\tau \, \partial_\tau \theta(x_j,\tau) +2\pi i(S-m).
\end{align}
If $S-m \not\in \mathbb Z$, the two states, $\ket{\psi_0}$ and $\ket{\psi_1}$, belong to different topological sectors.
When $S-m = p/q$ for coprime integers $p$ and $q$, there are $q$ orthogonal low-energy states,
$\ket{\psi_0}, \, \ket{\psi_1}, \, \cdots, \ket{\psi_{q-1}}$.
The low-energy state $\ket{\psi_n}$ for $n>0$ has the configuration,
\begin{align}
    \theta'(x,\tau) &= \theta(x,\tau) - \frac{2\pi a_0}{\beta L} n\tau.
\end{align}
Therefore, when $S-m \in \mathbb Z+ p/q$, the partition function at zero temperature is given by
\begin{align}
    Z &\approx  \sum_{n=0}^{q-1} \braket{\psi_n|e^{-\beta \mathcal H}|\psi_n}.
    \label{Z_p/q}
\end{align}

\subsection{Dual transformation to sine-Gordon theory}

\subsubsection{When $S-m \in \mathbb Z$}\label{app:S-m=int}

Here, we consider the simplest case of $S-m \in \mathbb Z$.
The $m\not=0$ case starts from the classical configuration \eqref{Omega_cl_plateau}.
Unlike the $m=0$ case, the topological excitation is the vortex rather than the meron.
Note that the path integral is defined under the (imaginary-)temporal boundary condition:
\begin{align}
    \theta(\tau+\beta,x) = \theta(\tau,x)  \mod 2\pi.
    \label{theta_bc_tau}
\end{align}
The LSM term \eqref{S_LSM} does not interfere with the boundary condition \eqref{theta_bc_tau} for $S-m \in \mathbb Z$ and is thus negligible.
For $S-m \in \mathbb Z$, the dual field theory is derived as follows.
The classical configuration,
\begin{align}
    n^x &= \sqrt{1-(m/S)^2}\, \cos \theta(x,\tau), \\
    n^y &= \sqrt{1-(m/S)^2} \, \sin \theta(x,\tau), \\
    a_0L^z &= m,
\end{align}
leads to the following action,
\begin{align}
    \mathcal S 
    &= \frac{v[1-(m/S)^2]}{2g} \int d\tau dx \, (\partial_\mu \theta)^2 - (\ln \zeta) \sum_{n=0}^{N_{\rm v}} \nu_n^2
    \notag \\
    &\qquad -i \pi (S-m) \sum_{n=0}^{N_{\rm v}} \nu_n
    \notag \\
    &= \frac{vK}{2\pi} \int d\tau dx \, (\partial_\mu \theta)^2 - (\ln \zeta) \sum_{n=0}^{N_{\rm v}} \nu_n^2
    \notag \\
    &\qquad -i \pi (S-m) \sum_{n=0}^{N_{\rm v}} \nu_n,
\end{align}
with the Luttinger parameter,
\begin{align}
    K &= \frac \pi g [1-(m/S)^2].
    \label{K_m}
\end{align}
$N_{\rm v}=0,1,2,\cdots$ is the number of vortices.
The Hubbard-Stratonovich transformation turns the partition function into
\begin{align}
     Z &= \sum_{N_{\rm v}} \int \mathcal D\theta \mathcal D J_\tau \mathcal DJ_x \,  \exp\biggl( -\frac{1}{2\pi K} \int d\tau dx\, {J_\mu}^2 
    \notag \\
    &\quad +\frac{i}{\pi} \int d\tau dx \,J_\mu \partial_\mu \theta + (\ln \zeta) \sum_{n=0}^{N_{\rm v}} \nu_n^2 +\mathcal S_{\rm BP}
    \biggr)
    \notag \\
    &= \sum_{N_{\rm v}=0}^\infty \int \mathcal D\phi\, e^{-\mathcal S_{\rm LSM}} \prod_{\tau,x} \exp\biggl(-\frac{1}{2\pi K}(\partial_\mu \phi)^2 + 2i\phi \nu_n
    \notag \\
    &\quad +(\ln \zeta) \nu_n^2 + \pi i(S-m) \nu_n \biggr).
    \label{Z_J_plateau}
\end{align}
Collecting the $N_{\rm v}=0,1$ terms, we obtain the following representation of the partition function,
\begin{align}
    Z &\approx \int \mathcal D\phi \prod_{\tau,x} \exp\biggl(-\frac{1}{2\pi K}(\partial_\mu \phi)^2\biggr)
    \notag \\
    &\qquad \times \biggl[1+ \zeta (e^{i[\pi (S-m)+2\phi]} + \mathrm{H.c.}) 
    \biggr]
    \notag \\
    &= \int \mathcal D\phi \prod_{\tau,x} \exp(-\mathcal S_{\rm dual}),
\end{align}
with the dual action
\begin{align}
    \mathcal S_{\rm dual}
    &=  \frac{v}{2\pi K} \int d\tau dx \, (\partial_\mu \phi)^2
    \notag \\
    &\qquad -2\zeta  \int d\tau dx \, \cos[\pi (S-m) + 2\phi]
    \notag \\
    &= \frac{v}{2\pi K} \int d\tau dx \, (\partial_\mu \phi)^2
    \notag \\
    &\qquad -2\zeta \cos[\pi(S-m)]  \int d\tau dx \, \cos(2\phi).
\end{align}
Though the dual action is qualitatively the same as Eq.~\eqref{Sz2phi_m=0}, they have two quantitative differences.
The first difference is the coupling constant of the cosine interaction, where $\Theta=2\pi S$ is replaced by $2\pi (S-m)$.
The other is the Luttinger parameter \eqref{K_m}.

Let us compile the translation dictionary from the spin to the boson.
We again start with $L^z$, which admits the following quantum fluctuation:
\begin{align}
    a_0L^z &= m + \frac{a_0}{gv}(\bm n \times \partial_t \bm n)^z
    \notag \\
    &= m + \frac{a_0K }{\pi v}\partial_t \theta
    \notag \\
    &= m + \frac{a_0}\pi \partial_x \phi.
\end{align}

Previously, we saw that the sine-Gordon theory at $m=0$ admits the staggered magnetic field $h_s$ as the imbalance of the fugacities of merons.
This time, however, the staggered field hardly affects the fugacities of vortices, because the vortices have no longitudinal $n^z$ component classically.
Instead, the staggered magnetic field is introduced to the action of the vortex through $m$.
The staggered field modifies $m$ to $m + (-1)^j \delta m$.
The staggered field doubles the unit-cell size of the spin-$S$ HAFM chain.
The unit cell contains two sites, $x_{2j'-1}$ and $x_{2j'}$.
The staggered magnetic field modifies the classical configuration to
\begin{align}
    \bm \Omega(x_{2j'-1}, \tau) &=
    \begin{pmatrix}
    - \sqrt{1- [(m-\delta m)/S]^2} \cos\theta_1(x_{2j'-1},\tau) \\
    -  \sqrt{1- [(m-\delta m)/S]^2}\sin \theta_1 (x_{2j'-1},\tau) \\
    (m-\delta m)/S
    \end{pmatrix}
    \label{Omega1_cl},
    \\
    \bm \Omega(x_{2j'}, \tau) &=
    \begin{pmatrix}
     \sqrt{1- [(m+\delta m)/S]^2} \cos\theta_2(x_{2j'},\tau) \\
    \sqrt{1- [(m+\delta m)/S]^2}\sin \theta_2 (x_{2j'},\tau) \\
    (m+\delta m)/S
    \end{pmatrix}.
    \label{Omega2_cl}
\end{align}
Doubling the unit-cell size also doubles the number of fields.
Note that $\theta_1+\theta_2$ is related to the global U(1) spin-rotation symmetry but $\theta_1-\theta_2$ is unrelated to any global symmetry.
Under such circumstance, the antisymmetric field, $\theta_1-\theta_2$, acquires the larger excitation gap than the symmetric field, $\theta_1+\theta_2$, does~\cite{oya}.
Integrating out the high-energy antisymmetric field imposes a  constraint, 
\begin{align}
    \theta_1(x,\tau) &= \theta_2(x,\tau) = \theta(x,\tau) \mod 2\pi.
\end{align}
This relation allows us to calculate the Berry phase in analogy with Eq.~\eqref{BP_tOmega}.
Namely, the staggered magnetization leads to the Berry phase:
\begin{widetext}
\begin{align}
    \mathcal S_{\rm BP}
    &= - iS\sum_{j'=1}^{L/2}\omega[\bm \Omega(x_{2j'},\tau)] -iS \sum_{j'=1}^{L/2} \biggl[2\biggl( 1-\frac{m+\delta m}{S} \biggr) \int_0^\beta d\tau \, \partial_\tau \theta_1(x_{2j'-1},\tau) - \omega[\bm \Omega(x_{2j'-1},\tau) ]
    \biggr]
    \notag \\
    &= -iS\sum_{j=1}^L (-1)^j \omega[\bm \Omega(x_j,\tau)] -2i(S-m-\delta m) \sum_{j'=1}^{L/2} \int_0^\beta d\tau \, \partial_\tau \theta(x_{2j'-1},\tau)
    \notag \\
    &= \pi i(S-m-\delta m) Q_{\rm v} -i(S-m) \int_0^L \frac{dx}{a_0} \int_0^\beta d\tau \, \partial_\tau \theta(x,\tau).
    \label{BP_vortex_dm}
\end{align}
\end{widetext}
The constant shift $S-m \to S-m -\delta m$ of the coefficient of $Q_{\rm v}$ affects the dual action as follows.
\begin{align}
    \mathcal S_{\rm dual}
    &= \frac v{2\pi K} \int d\tau dx \, (\partial_\mu \phi)^2
    \notag \\
    &\quad -2\zeta \int d\tau dx \, \cos[\pi (S-m-\delta m) -2\phi]
    \notag \\
    &\approx \frac{v}{2\pi K} \int d\tau dx \, (\partial_\mu \phi)^2
    \notag \\
    &\quad -2\zeta \int d\tau dx \, \cos[\pi (S-m)] \cos(2\phi)
    \notag \\
    &\quad +2\zeta \delta m \int d\tau dx \, \cos[\pi (S-m)]\sin(2\phi).
\end{align}
The last term implies that $n^z$ is given by
\begin{align}
    n^z &\propto (-1)^j \cos[\pi (S-m)] \sin(2\phi).
    \label{nz_S-m_int}
\end{align}
$L^x$, $L^y$ are obtained similarly to Eqs.~\eqref{Lx2phi} and \eqref{Ly2phi}.
In short, we obtain
\begin{align}
    S_j^z &= m + \frac{a_0}{\pi} \partial_x \phi + (-1)^j a_1 \cos[\pi (S-m)] \sin(2\phi),
    \label{Sz2phi_S-m_int}
    \\
    S_j^+ &= e^{i\theta} \bigl[(-1)^j + b_1 \cos[\pi (S-m)] \sin(2\phi) \bigr],
    \label{S+2phi_S-m_int}
\end{align}
with nonuniversal constants, $a_1, b_1 \in \mathbb R$.
The semiclassical bosonization formulas \eqref{Sz2phi_S-m_int} and \eqref{S+2phi_S-m_int} are identical to the $m=0$ ones~\eqref{Sz2phi_m=0} and \eqref{S+2phi_m=0} by replacing $\sin(\pi S)$ with $\cos[\pi (S-m)]$.
The bosonization formulas \eqref{Sz2phi_S-m_int} and \eqref{S+2phi_S-m_int} indicate that the one-site translation $T_1: \bm S_j \to \bm S_{j+1}$ and the site-centered inversion $\mathcal I_s: \bm S_j \to \bm S_{L-j}$ act on $\phi$ and $\theta$ as
\begin{align}
    T_1&:\,  \phi(x,\tau)  \to \phi(x,\tau) + \frac \pi 2,
    \label{T1_S-m=1/2_phi}
    \\
    T_1&:\, \theta(x,\tau) \to \theta(x,\tau) + \pi,
\end{align}
and 
\begin{align}
    \mathcal I_s&:\, \phi(x,\tau) \to -\phi(L-x,\tau) + \frac \pi 2, \\
    \mathcal I_s&:\, \theta(x,\tau) \to \theta(L-x,\tau) + \pi.
\end{align}
Since the bond-centered inversion $\mathcal I_b = T_1\mathcal I_s$ keeps the spin-rotation symmetries and inverts $\phi(x,\tau) \to -\phi(L-x,\tau)$, the dimer order parameter $(-1)^j\bm S_j \cdot \bm S_{j+1}$ is again bosonized as
\begin{align}
    (-1)^j \bm S_j \cdot \bm S_{j+1} &= d \cos(2\phi) + \cdots.
\end{align}

\subsubsection{When $S-m\in \mathbb Z+p/q$}

When $S-m$ has a decimal part, we need to include the LSM-twisted states [Eq.~\eqref{Z_p/q}].
If we include the vortices with $\nu=\pm 1$ only, the partition function would become
\begin{align}
    Z &\approx \int \mathcal D\phi \prod_{\tau,x} \exp\biggl( -\frac 1{2\pi K}(\partial_\mu \phi)^2 \biggr)
    \notag \\
    &\quad \times \sum_{n=0}^{q-1} \biggl[ 1+ e^{i\frac{2\pi n}q} \zeta ( e^{i[\pi (S-m) - 2\phi]} + e^{-i[\pi (S-m)-2\phi]})
    \biggr]
    \notag \\
    &=\int \mathcal D\phi \prod_{\tau,x} \exp\biggl( -\frac 1{2\pi K}(\partial_\mu \phi)^2 \biggr),
\end{align}
because $\sum_{n=0}^q e^{i(2\pi n/q)} = 0$.
The LSM term thus forbids the single-vortex excitation with $\nu=\pm 1$.
Likewise, it forbids those with $\nu\not=0 \mod q$.
Therefore, vortices can be excited only when $\nu_n \in q\mathbb Z$.
The dual action is then given by
\begin{align}
    \mathcal S_{\rm dual}
    &= \frac{v}{2\pi K} \int d\tau dx \, (\partial_\mu \phi)^2
    \notag \\
    &\qquad - 2\zeta \int d\tau dx \, \cos[\pi q(S-m) -2q\phi] \notag \\
    &= \frac v{2\pi K} \int d\tau dx \, (\partial_\mu \phi)^2 
    \notag \\
    &\qquad -2\zeta \cos[\pi q(S-m)] \int d\tau dx \, \cos(2q\phi).
\end{align}
This field theory is perfectly consistent with the one discussed by Oshikawa, Yamanaka, and Affleck~\cite{oya}.

\subsubsection{When $S-m \in \mathbb Z+ 1/2$}\label{app:S-m=1/2}

In the main text, the semiclassical bosonization formulas are required for the $S-m\in\mathbb Z+1/2$ case.
In what follows, we limit ourselves to the $S-m\in \mathbb Z+1/2$ case.

To derive the semiclassical bosonization formulas, we need to investigate the response of the dual field theory to the external staggered magnetic field.
There is a low-energy state, $\ket{\psi'_0}$, that lives in a different topological sector from the ground state $\ket{\psi_0}$ for $\delta m=0$.
Previously, we defined $\ket{\psi'_0} = \ket{\psi_1}$ by twisting the $\theta$ field [Eq.~\eqref{theta'_LSM}].
When $Q_{\rm v}=\pm 1$ and $S-m \in\mathbb Z + 1/2$, 
we can instead employ another, more convenient, example of a low-energy state $\ket{\psi'_0}$, that is,
\begin{align}
    \ket{\psi'_0} &= \mathcal I_b \ket{\psi_0}.
\end{align}
The bond-centered inversion flips the sign of the vorticity density,
\begin{align}
    \mathcal I_b:\, \nu_n \to - \nu_n.
    \label{I_b_nu}
\end{align}
Figure~\ref{fig:sq} gives a schematic understanding of the relation \eqref{I_b_nu}.

The $2i\phi \nu_n$ term of Eq.~\eqref{Z_J_plateau} is invariant under $\mathcal I_b$ because this term is originally $iJ_\mu \partial_\mu \theta_{\rm v}/\pi =- K(\partial_\mu \theta_{\rm v})^2/\pi$, which is apparently $\mathcal I_b$-invariant.
To compensate this sign and make this term $\mathcal I_b$-invariant, the $\phi$ field must transform as
\begin{align}
    \mathcal I_b:\, \phi(x) \to - \phi(L-x) \mod 2\pi .
    \label{Ib_J4_chain}
\end{align}
The relation \eqref{I_b_nu} results in an interesting fact that $\ket{\psi_0}$ and $\ket{\psi'_0}$ belong to different topological sectors because 
\begin{align}
    \mathcal I_b \mathcal S_{\rm BP} \mathcal I_b^{-1}
    &= -iS\sum_{j=1}^L (-1)^j \omega[\tilde{\bm \Omega}(x_{L+1-j},\tau)]
    \notag \\
    &\quad
    -2i(S-m) \sum_{j'=1}^{L/2} \int_0^\beta d\tau\, \partial_\tau \theta(x_{L+1-(2j'-1)},\tau)
    \notag \\
    &= iS \sum_{j=1}^L (-1)^j \omega[\tilde{\bm \Omega}(x_j,\tau)] \notag \\
    &\quad -2i(S-m) \sum_{j'=1}^{L/2} \int_0^\beta d\tau \, \partial_\tau \theta(x_{2j'},\tau)
    \notag \\
    &=\mathcal S_{\rm BP} -2\pi i(S-m) Q_{\rm v}.
\end{align}
If $Q_{\rm v}=1$ and $S-m\in \mathbb Z+1/2$, the following relation holds:
\begin{align}
    \mathcal I_b \mathcal S_{\rm BP} \mathcal I_b^{-1} - \mathcal S_{\rm BP} &= \pi i \mod 2\pi i.
\end{align}

The two states $\ket{\psi_0}$ and $\mathcal I_b\ket{\psi_0}$ are thus orthogonal low-energy states, equally contributing to the partition function,
\begin{align}
    Z &=  \braket{\psi_0|e^{-\beta \mathcal H}|\psi_0} +  \braket{\psi'_0|e^{-\beta \mathcal H}|\psi'_0} 
    \notag \\
    &= \braket{\psi_0|e^{-\beta \mathcal H}|\psi_0} +  \braket{\psi_0|\mathcal I_b^{-1} e^{-\beta \mathcal H}\mathcal I_b|\psi_0} .
\end{align}
When collecting the $N_{\rm v}=0, 1$ terms, we find
\begin{align}
    Z
    &\approx \int \mathcal D \phi \prod_{\tau,x} \exp\biggl(-\frac v{2\pi K} (\partial_\mu \phi)^2 \biggr)
    \notag \\
    &\quad \times \biggl[1 + \zeta (e^{i[\pi (S-m) -2\phi]} + \mathrm{H.c.} )
    \notag \\
    &\quad - \zeta (e^{i[\pi (S-m) -2\phi]} +  \mathrm{H.c.})
    \biggr]
    \notag \\
    &= \int \mathcal D \phi \prod_{\tau,x} \exp\biggl(-\frac v{2\pi K} (\partial_\mu \phi)^2 \biggr).
\end{align}
The $N_{\rm v}=2$ term needs to be included to the action in order to generate the most relevant interaction.
The dual action thus becomes
\begin{align}
    \mathcal S_{\rm dual} &= \frac{v}{2\pi K} \int d\tau dx \, (\partial_\mu \phi)^2
    \notag \\
    &\qquad -2\zeta^2 \cos[2\pi (S-m)] \int d\tau dx \, \cos(4\phi).
    \label{S_dual_S-m=1/2}
\end{align}
If the ground state of the spin-1 HAFM chain for $S-m=1/2$ is gapped, it must be doubly degenerate by spontaneously breaking the $\phi\to \phi+\frac{\pi}{2}$ symmetry.
The $\phi\to \phi+\frac{\pi}{2}$ symmetry is highly likely to be the one-site translation symmetry.
If so, the site-centered inversion symmetry $\mathcal I_s = \mathcal I_b T_1$ acts on $\phi(x,\tau)$ as
\begin{align}
    \mathcal I_s:\, \phi(x) \to - \phi(L-x) + \frac{\pi}{2}\mod 2\pi.
\end{align}
Like $\mathcal I_b$, the site-centered inversion $\mathcal I_s=\mathcal I_bT_1$ flips the sign of $\nu_n\to -\nu_n$ because $T_1$ keeps $\nu_n$.
An extra phase arises from the coupling $2i\phi \nu_n$ in Eq.~\eqref{Z_J_plateau}.
In fact,
\begin{align}
    \mathcal I_s (2i\phi \nu_n) \mathcal I_s^{-1}
    &= 2i\phi \nu_n + \pi i \nu_n,
\end{align}
contains the phase $\pi i$ for $\nu_n=\pm 1$.
Thus, the state $\mathcal I_s\ket{\psi_0}$ also belongs to the different topological sector from  that $\ket{\psi_0}$ lives in.
Just like we did for $\mathcal I_b$, we can confirm that the $\nu_n=\pm 1$ contributions to the dual action $\mathcal S_{\rm dual}$ are canceled between the two low-energy states, $\ket{\psi_0}$ and $\mathcal I_s\ket{\psi_0}$.

Let us apply the staggered magnetic field, $h_s\sum_j (-1)^j S_j^z$, to the spin-$S$ HAFM chain.
The staggered magnetic field keeps the $\mathcal I_s$ symmetry but breaks the $\mathcal I_b$ symmetry.
The Berry phase \eqref{BP_vortex_dm} modified by the staggered magnetic field leads to
\begin{align}
    Z 
    &\approx \int \mathcal D\phi \, \prod_{\tau,x} \exp\biggl( -\frac{1}{2\pi K} (\partial_\mu \phi)^2 \biggr)
    \notag \\
    &\quad \times \biggl[ 1+ \zeta (e^{i[\pi (S-m-\delta m) +2\phi]} + \mathrm{H.c.} )
    \notag \\
    &\quad + e^{\pi i} \zeta e^{i[-\pi (S-m-\delta m) +2\phi]} + \mathrm{H.c.} )
    \biggr]
    \notag \\
    &=\int \mathcal D\phi \, \prod_{\tau,x} \exp\biggl( -\frac{1}{2\pi K} (\partial_\mu \phi)^2 \biggr)
    \notag \\
    &\quad \times \biggl[ 1- 4\zeta\sin[\pi (S-m-\delta m)]\sin(2\phi)
    \biggr].
\end{align}
We thus obtain the dual action
\begin{align}
    \mathcal S_{\rm dual}
    &= \frac{v}{2\pi K} \int d\tau dx \, (\partial_\mu\phi)^2 
    \notag \\
    &\qquad +4\pi \zeta \delta m \int d\tau dx \, \cos[\pi(S-m)] \sin(2\phi).
\end{align}
This dual action implies the following representation of the staggered magnetization,
\begin{align}
    n^z &\propto \cos[\pi (S-m)] \sin(2\phi),
\end{align}
leading to the following semiclassical bosonization formulas \eqref{Sz2phi_S-m_int} and \eqref{S+2phi_S-m_int}.
The bond alternation $(-1)^j\bm S_j \cdot \bm S_{j+1}$ is bosonized as
\begin{align}
    (-1)\bm S_j \cdot \bm S_{j+1} &= d\cos(2\phi)+\cdots.
\end{align}

\section{$J_4$ interaction}
\label{app:J4}

In the main text, we deal with an interaction,
\begin{align}
    \mathcal V_4
    &=J_4 \sum_{j=1}^{L/2}(-1)^j\bm S_{2j-1}\cdot \bm S_{2j}
    \notag \\
    &= -J_4\sum_{j=1}^{L-1} \sin\biggl(\frac{\pi j}{2}\biggr)\bm S_j \cdot \bm S_{j+1}.
    \label{J_4}
\end{align}
Let us include $\mathcal V_4$ into the low-energy effective field theory of the spin-1 chain perturbatively.
Here, the full Hamiltonian is
\begin{align}
    \mathcal H_4 &= \mathcal H_0 + \mathcal V_4, \\
    \mathcal H_0 &= J\sum_{j'=1}^{L/2} (\bm S_{2j'-1}\cdot \bm S_{2j'}+\alpha \bm S_{2j'}\cdot \bm S_{2j'+1}).
\end{align}
This perturbative expansion can be systematically performed~\cite{furuya_screw}.
Let $P$ be a projection operator to the low-energy subspace of the unperturbed model $\mathcal H_0$.
Acting on the unperturbed ground state, the interaction \eqref{J_4} generates an excitation with $q=\pi/2a_0$, which is almost at the top of the single-band excitation band, that is, $P\mathcal V_4 P = 0$.
Hence, the interaction \eqref{J_4} does not affect the ground state and low-energy physics up to the first-order of the perturbative expansion.
The leading contribution comes from the second-order perturbation,
\begin{align}
    P \mathcal V_4 \frac{1}{E_0 - \mathcal H_0} Q\mathcal V_4 P,
\end{align}
where $Q=1-P$.
The second-order perturbation gives rise to various interaction.
The most relevant one is the biquadratic interaction,
\begin{align}
    &-\lambda_4 \sum_{j=1}^L \sin^2\biggl(\frac{\pi j}{2}\biggr)(\bm S_j \cdot \bm S_{j+1})^2
    \notag \\
    &= - \frac{\lambda_4}{2}\sum_{j=1}^L \bigl\{1-(-1)^j\bigr\}(\bm S_j \cdot \bm S_{j+1})^2
\end{align}
with $\lambda_4\propto {J_4}^2/J$ is a positive constant.

Using the semiclassical bosonization formulas for $S-m=1/2$ (Appendix~\ref{app:S-m=1/2}), we obtain 

\begin{align}
    \sum_{j=1}^L (\bm S_j\cdot \bm S_{j+1})^2 &= d_u \int_0^Ldx \, \cos^2(2\phi) + \cdots
    \notag \\
    &= \frac{d_u}{2}\int_0^L dx \,  \cos(4\phi) + \cdots
\end{align}
with $d_u\propto d^2 >0$.
On the other hand, the staggered part contains $\sin(2\phi)$,
\begin{align}
    \sum_{j=1}^L (-1)^j (\bm S_j \cdot \bm S_{j+1})^2 = d_s \int_0^L dx \, \cos(2\phi) + \cdots,
\end{align}
with a constant $d_s=m^2 d$.
The constant $d$ relates the dimerization with $\phi$: $(-1)^j \bm S_j \cdot \bm S_{j+1} =d \cos(2\phi)+\cdots$~\cite{takayoshi_coeff,hikihara_coeff_dimer}.
We can set the constant $d>0$ without loss of generality.
The effective Hamiltonian of the spin-1 chain \eqref{H_4} of the main text on the $1/2$ plateau is 
\begin{align}
    \mathcal H_4
    &=  \frac{v}{2\pi K}\int_0^L dx \, (\partial_\mu \phi)^2+ (g_2(J_4)-g_{2c}) \int_0^L dx \, \cos(2\phi) 
    \notag \\
    &\quad +g_4(J_4) \int_0^L dx \, \cos(4\phi),
    \label{H_4_phi}
\end{align}
with $g_{2c} \propto J(1-\alpha)$, $g_2(J_4)=\lambda_4 d_s/2>0$, and $g_4(J_4)=-\lambda_4 d_u/2 < 0$.
This field theory is called the double sine-Gordon theory.
Note that $\exp(\pm 2i\phi)$ are the most relevant vertex operators in accordance with the compactification relation, $\phi\sim \phi+\pi$ for $S-m \in \mathbb Z$.
When $g_2(J_4)-g_{2c}\not=0$, the $\cos(4\phi)$ interaction is negligible since it is less relevant than $\cos(2\phi)$ even if $\cos(4\phi)$ is relevant.
The sign of $g_2(J_4)-g_{2c}$ determines the ground state and the edge magnetization.
The $\phi$ field is locked to $\bar\phi=0$ (i.e. $\mathcal P=0$) for $g_{2}(J_4)-g_{2c}<0$ and to $\bar\phi=\pm \pi/2$ (i.e. $\mathcal P=\pm 1/2$) for $g_2(J_4)-g_{2c}>0$.
The quantum phase transition at $g_2(J_4)-g_{2c}=0$ is the second order if $\cos(4\phi)$ is irrelevant and otherwise the first order.

The quantum phase transition is likely to be the second order from the $J_4$ dependence of the excitation gap [Fig.~\ref{fig:gap_ee}~(a)].
The quantum critical behavior is also supported by the site-dependent entanglement entropy [Fig.~\ref{fig:gap_ee}~(b)].
If $\cos(4\phi)$ is irrelevant, the central charge at $g_2(J_4)-g_{2c}=0$ is exactly $c=1$.
We fitted the numerical data by Eq.~\eqref{S_EE} by regarding $a_s$ and $c$ of Eq.~\eqref{S_EE} as fitting parameters [Fig.~\ref{fig:gap_ee}~(b)].
We obtained $(a_s,c) = (0.52,\, 0.94)$, consistent with the quantum field theory \eqref{H_4_phi} with irrelevant $\cos(4\phi)$.

\begin{figure}
    \centering
    \includegraphics[bb = 0 0 864 504, width=\linewidth]{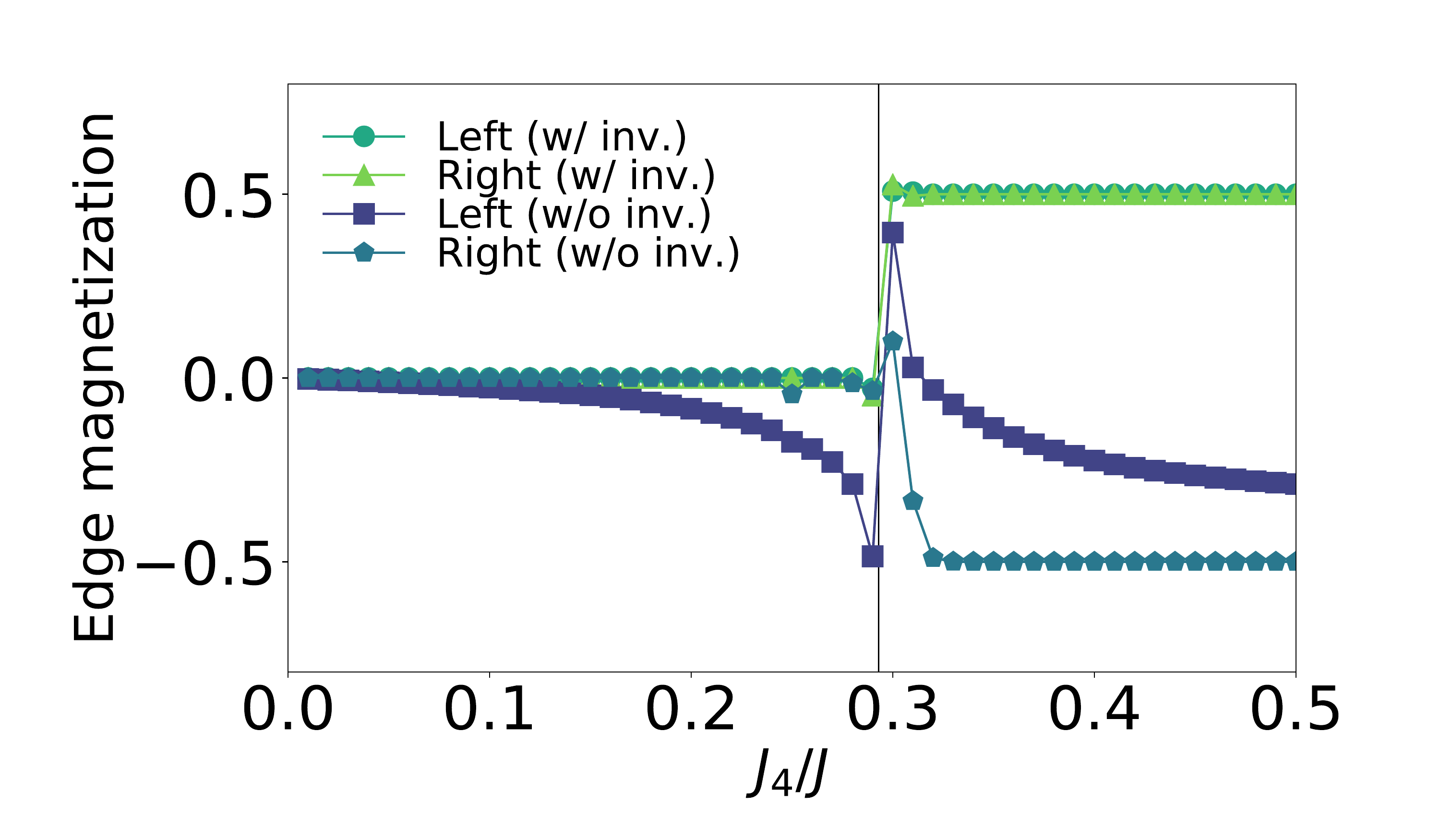}
    \caption{Comparison of edge magnaetizations $M_{\rm right}^z$ and $M_{\rm left}^z$ for $L=242$ (circles and triangles) and for $L=240$ (squares and pentagons).
    The other parameters are fixed to $h_u/J=1.5$ and $\alpha=0.2$.
    The former case has the $I_b$ symmetry and the latter has not.
    The inversion symmetry quantizes the edge magnetizations except in the vicinity of the quantum critical point $J_4/J=0.295$.
    Without the inversion symmetry, the edge magnetization shows continuous changes even away from the quantum critical point.
    }
    \label{fig:edge_wo_inversion}
\end{figure}

Despite the quantum critical behaviors of the excitation gap and the entanglement entropy, the order parameters show discontinuous behaviors because of the inversion symmetry.
Figure~\ref{fig:edge_wo_inversion} shows the edge magnetizations with and without the $\mathcal I_b$ symmetry.
The $\mathcal I_b$ symmetry imposes a strong constraint on the edge magnetization as the bulk polarization.
Recall that $U$ is given by Eq.~\eqref{U_def_plateau}.
The exact $\mathcal I_b$ symmetry of the model \eqref{H_4} leads to
\begin{align}
    \braket{U}
    &= \braket{\mathcal I_b U \mathcal I_b^{-1}}
    \notag \\
    &= \biggl\langle \exp \biggl(i\frac{2\pi}{L}\sum_{j=1}^Lj C_{L+1-j} \biggr) \biggr\rangle
    \notag \\
    &=\biggl\langle \exp \biggl(i\frac{2\pi}{L}\sum_{j=1}^L(L+1-j) C_{j} \biggr) \biggr\rangle
    \notag \\
    &= \braket{U^\dag} \exp\biggl(2\pi i \biggl(1+\frac 1L\biggr) \sum_{j=1}^L \braket{C_j^z} \biggr)
    \label{U_Ib}
\end{align}
The charge neutrality condition, $\sum_{j=1}^L \braket{C_j}=0$, leads to $\braket{U}=\braket{U^\dag}$.
Accordingly, if $\braket{U}\not=0$,
\begin{align}
    \mathcal P &= \frac{1}{2\pi} \im\ln \braket{U} = 0 \text{ or } \frac 12 \mod 1.
    \label{edge_Ib}
\end{align}
The condition $\braket{U}\not=0$ is equivalent to the locking of $\phi$.
$\braket{U}=0$ holds at the quantum critical point where the $\phi$ field is gapless, where the edge polarization is ill-defined.

In the absence of the $\mathcal I_b$ symmetry, the edge magnetization is not necessarily quantized.
Indeed, Fig.~\ref{fig:edge_wo_inversion} shows that $M_{\rm left}^z$ depends on $J_4$ continuously except for the quantum critical point $J_4=J_{4c}$.
By contrast, the other edge $M_{\rm right}^z$ is well quantized.
This asymmetric behavior of the quantization can be understood as the modification of the boundary condition.
The spin chain \eqref{H_4} has the exact $\mathcal I_b$ symmetry with the OBC for $L=2 \mod 4$.
By adding two sites to one edge of the chain, say, the left edge, we can make $L=0 \mod 4$ and violate the $\mathcal I_b$ symmetry.
We can expect that such addition of sites hardly modifies the bulk Hamiltonian and generate a symmetry-breaking potential localized at the left edge of the chain.
The most relevant interaction is the staggered magnetic field, $h_B\sin(2\phi(0,\tau))$.
This boundary staggered field alters the boundary condition on the left edge from $\phi(x=0)=0$ to
\begin{align}
    \phi(x=0,\tau) = \Phi,
\end{align}
with a constant $\Phi$.
This modification of the boundary condition explains the continuous change of the edge magnetization (Fig.~\ref{fig:edge_wo_inversion}) in the absence of the $\mathcal I_b$ symmetry.

\begin{figure}
    \centering
    \includegraphics[bb = 0 0 900 1000, width=\linewidth]{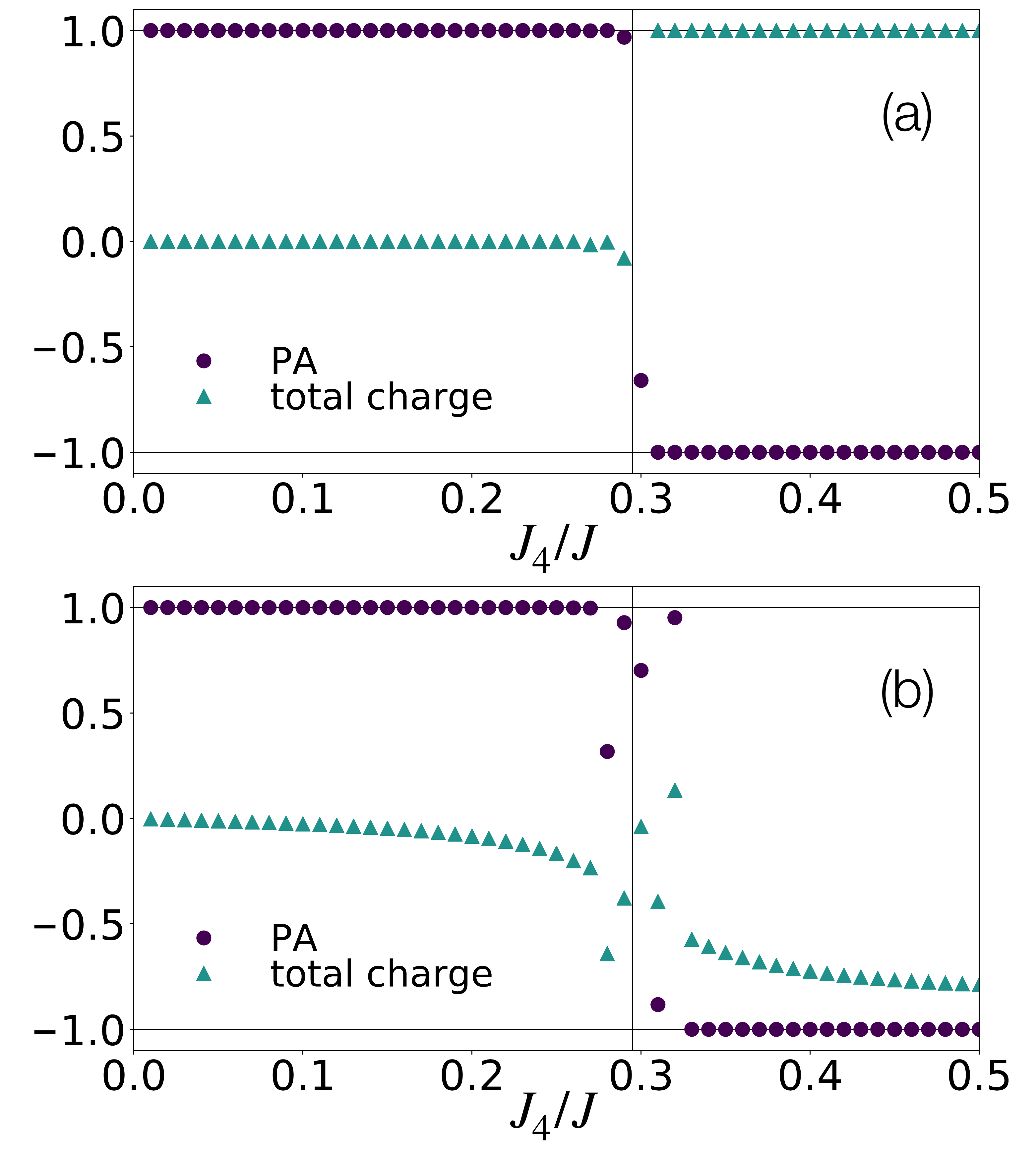}
    \caption{$J_4$ dependence of polarization amplitude $\braket{U}$ (circles) and total charge $\sum_{j=1}^L\braket{C_j}$ (triangles) for $h_u/J=1.5$, $\alpha=0.2$ with system size (a) $L=242$ and (b) $L=240$.
    (a) When the $\mathcal I_b$ symmetry is present, the polarization amplitude and the total charge show abrupt changes at $J_4/J\approx0.295$ thanks to the $\mathcal I_b$ symmetry.
    (b) On the other hand, when the $\mathcal I_b$ symmetry is absent, the total charge changes continuously though the accuracy of the numerical data is extremely lowered in the vicinity of the quantum critical point $J_4/J\approx 0.295$. Note that apart from $\braket{U}$ and the total charge, the other thermodynamic quantities behave similarly regardless of the presence of the absence of the $\mathcal I_b$ symmetry.
    }
    \label{fig:twist_charge}
\end{figure}

The edge magnetization as the bulk polarization \eqref{edge_Ib} works as an order parameter to distinguish the $\bar\phi=0$ phase ($\mathcal P=0$) and the $\bar\phi=\pi/2$ phase ($\mathcal P=1/2 \mod 1$).
Likewise, the polarization amplitude $\braket{U}$ can also be regarded as the order parameter.
Figure~\ref{fig:twist_charge} shows the $J_4$ dependence of the polarization amplitude $\braket{U}$ and the total charge, $\sum_{j=1}^L\braket{C_j}$.
The abrupt jump of the order parameter \eqref{edge_Ib} is due to the $\mathcal I_b$ symmetry.
The polarization amplitude $\braket{U}$ also jumps unlike the bond-alternating spin-$S$ chains at zero magnetic fields~\cite{nakamura-todo}, though the polarization amplitude is not necessarily quantized as $\braket{U}=\pm 1$.
If the excitation gap closes at the transition point, the polarization amplitude will change continuously and cross zero at the transition point.
Therefore, the observed jump of Fig.~\ref{fig:twist_charge} implies that the quantum critical regime is too narrow to observe such a continuous change of $\braket{U}$.

The total charge can also be seen as an order parameter to distinguish the two phases of concern [Fig.~\ref{fig:twist_charge}~(a)].
The charge neutrality is weakly broken for $J_4>J_{4c}$.
Nevertheless, since the discrepancy of the charge is precisely one, the quantum phase remains on the $1/2$ magnetization plateau in the $L\to + \infty$ limit, $M/M_s = \frac 12 + \frac 1L \to \frac 12$.

The $J_4$ dependence of the total charge becomes continuous in the absence of the $\mathcal I_b$ symmetry [Fig.~\ref{fig:twist_charge}~(b)].
Besides, the total charge tends to $-1$ in the $J_4/J$ limit.
The sign change compared to the $I_b$-symmetric case remains obscure at this stage.
Except for this continuous decrease of the total charge, the numerical results and the field-theoretical analyses are consistent.

\section{Union-jack strip}
\label{app:unionjack}

The effective field theory \eqref{H_UJ_phi} of the union-jack strip 
\begin{align}
\mathcal{H}_{\rm UJ}
    &= J_1\sum_{j=1}^L\sum_{n=1}^3 \bm S_{j,n} \cdot \bm S_{j+1,n} + J_1 \sum_{j=1}^{L}\bm S_{j,2}\cdot (\bm S_{j,1}+\bm S_{j,3})
    \notag \\
    &+J_2\sum_{j=2}^{L-1}\bm S_{j,2}\cdot (\bm S_{j-1,1}+\bm S_{j-1,3} +\bm S_{j+1,1}+\bm S_{j+1,3})
    \notag \\
    &-h_u \sum_{j=1}^L \sum_{n=1}^3 S_{j,n}^z,
    \label{H_UJ_hu}
\end{align}
on the $1/3$ plateau is derived similarly to that in Appendix~\ref{app:m>0}.
The infinitesimal uniform magnetic field $h_u>0$ is imposed to break the $S^z\to-S^z$ symmetry.
We consider the classical configuration,
\begin{align}
    \bm\Omega_n(x_j,\tau) &=
    \begin{pmatrix}
    (-1)^{j+n} \sqrt{1-(m/S)^2}\cos\theta_n(x_j,\tau) \\
    (-1)^{j+n} \sqrt{1-(m/S)^2}\sin\theta_n(x_j,\tau) \\
    m/S
    \end{pmatrix},
\end{align}
for each leg.
The index $n=1,2,3$ denotes the $n$th leg.
The interleg interaction leads to
\begin{align}
    \theta_1(x,\tau) &=\theta_2(x,\tau)  = \theta_3(x,\tau),
\end{align}
modulo $2\pi$ in the ground state.
The Berry phase is thus given by
\begin{align}
    \mathcal S_{\rm BP}
    &=3\pi i (S-m) Q_{\rm v} -3i(S-m) \int_0^L \frac{dx}{a_0} \int_0^\beta d\tau \, \partial_\tau \theta(x,\tau),
\end{align}
where $\theta=(\theta_1+\theta_2+\theta_3)/3$.
Since $3(S-m) = 1$ for the spin-$1/2$ union-jack strip on the $1/3$ plateau, the dual theory is derived similarly to Appendix~\ref{app:S-m=int}.
\begin{align}
    \mathcal S_{\rm dual}
    &= \frac{v}{2\pi K}\int d\tau dx \, (\partial_\mu \phi)^2 \notag \\
    &\qquad -2\zeta \cos[3\pi(S-m)] \int d\tau dx \, \cos(2\phi),
\end{align}
with $\phi=\phi_1+\phi_2+\phi_3$.
In this formulation, the long-range antiferromagnetic order results from the locking of $\phi$.
The semiclassical bosonization formulas on the $1/3$ plateau differ from Eqs.~\eqref{Sz2phi_S-m_int} and \eqref{S+2phi_S-m_int} because the staggered magnetic field,
\begin{align}
    h_s\sum_{j=1}^L\sum_{n=1}^3 (-1)^{j+n}S_{j,n}^z,
\end{align}
keeps all the symmetries of the model \eqref{H_UJ_hu}.
This symmetry implies the bosonization formula,
\begin{align}
    (-1)^{j+n} S_{j,n}^z \propto \cos(2\phi).
    \label{nz_uj}
\end{align}

\bibliography{ref.bib}

\end{document}